\documentclass[pre,10pt,twocolumn,superscriptaddress,showpacs]{revtex4-1}
\usepackage{amsmath}
\usepackage{latexsym}
\usepackage{amssymb}
\usepackage{mathtools}
\usepackage{braket}
\usepackage{graphicx}
\usepackage[font=Large,caption=false]{subfig}
\usepackage{wrapfig}
\usepackage{dsfont}
\usepackage{etoolbox}
\usepackage[colorlinks=true, citecolor=blue, urlcolor=blue]{hyperref}
\usepackage{float}
\usepackage{amsfonts}

\newcommand{\be}{\begin{equation}}
\newcommand{\ee}{\end{equation}}
\newcommand{\ba}{\begin{eqnarray}}
\newcommand{\ea}{\end{eqnarray}}

\newcommand{\non}{\nonumber}

\DeclareMathOperator{\sinc}{sinc}
\begin{document}
\title{A quantum enhanced finite-time Otto cycle}

\author{Arpan Das}
\email{arpandas@imsc.res.in}
\affiliation{Optics and Quantum Information Group, The Institute of Mathematical Sciences, CIT Campus, Taramani, Chennai 600113, India.}

\author{Victor Mukherjee}
\email{mukherjeev@iiserbpr.ac.in}
\affiliation{Department of Physical Sciences, IISER Berhampur, Berhampur 760010, India}

\begin{abstract}
We use fast periodic control to realize finite-time Otto cycles exhibiting quantum advantage. Such periodic modulation of the working medium - bath interaction Hamiltonian during the thermalization strokes can give rise to non-Markovian anti-Zeno dynamics, and corresponding reduction in the thermalization times. Faster thermalization can in turn significantly enhance the power output in engines, or equivalently, the rate of refrigeration in refrigerators. This improvement in performance of dynamically controlled Otto thermal machines arises due to the time-energy uncertainty relation of quantum mechanics.
\end{abstract}

\maketitle

\section{Introduction}
The recent experimental advances in control of systems in the quantum regime \cite{golter16optomechanical, accanto17rapid, perreault17quantum, rossi18measurement}, have in part led to the current extensive interest in theoretical \cite{giovannetti11advances, kurizki15quantum} and experimental \cite{brantut13a, bernien17probing, zhang17observation, klatzow19experimental, peterson19experimental} studies of quantum technologies. One of the fundamental aspects of quantum technologies involve thermodynamics in the quantum regime \cite{kosloff13quantum, klimovsky15thermodynamics, vinjanampathy16quantum, alicki18introduction, binder18book}, and the related studies of engines and refrigerators \cite{alicki79the, klimovsky13minimal, brantut13a, alicki14quantum, rossnage14nanoscale, kosloff17the, ghosh17catalysis, campisi16the, klatzow19experimental, peterson19experimental, chen19an, hartmann19many, revathy20universal,alejandro,  jinfuchen1, jinfuchen2}, quantum batteries \cite{campaioli17enhancing, marcello1, marcello2, marcello3} and quantum 
probes \cite{correa15individual, correa16low, zwick17quantum, mukherjee19enhanced, bhattacharjee20quantum}. A major challenge in the field of quantum thermodynamics is to design optimally performing quantum thermal machines, which can operate with maximum efficiency, power, or refrigeration \cite{erdman19maximum}. Naturally, a question arises - can quantum effects boost the performance 
of these quantum machines \cite{harrow17quantum}? Recent studies have indeed shown the possibility of harnessing quantum effects to achieve quantum enhancement in quantum devices, for example in the context of quantum computing \cite{boixo18characterizing},  in quantum thermal machines over many cycles \cite{watanabe17quantum},  in interacting many-body quantum thermal machines in presence of non-adiabatic dynamics \cite{campo16quantum}, through collective coherent coupling to baths  \cite{niedenzu18cooperative, kloc19collective}, as well as experimentally, in presence of coherence \cite{klatzow19experimental}.

A relatively less explored area, which can prove to be highly beneficial for improving the performance of quantum technologies, is quantum machines exhibiting non-Markovian dynamics \cite{abiuso19non, mukherjee20anti, camati20employing}. Studies of quantum thermal machines in general involve analysis of quantum systems coupled to dissipative environments. Quantum technologies based on open quantum systems, undergoing Markovian dynamics \cite{breuer02}, have been studied extensively in the literature \cite{uzdin15equivalence, kosloff17the, niedenzu18cooperative, binder18book}. Yet,  Markovian approximation may become invalid, for example, in the presence of strong system-bath coupling, or small bath-correlation times, in which case, going beyond the Markovian approximation becomes essential \cite{chruscinski10non, klimovsky13work, rivas14quantum, gil16quantum, nahar19preparations}. However, several open questions remain regarding the thermodynamics of quantum systems undergoing non-Markovian dynamics, and the conditions under which non-Markovianity can prove to be advantageous for engineering quantum technologies \cite{mukherjee15efficiency, uzdin16quantum, thomas18thermodynamics, pezzutto19an}.\\\\ 
Here we show the possibility of achieving quantum advantage in quantum machines undergoing non-Markovian dynamics; we consider an Otto cycle, in presence of a working medium (WM) subjected to fast periodic modulations, in the form of rapid coupling / decoupling of the WM with the thermal baths during the thermalizing strokes. Modulations of the WM-bath interaction Hamiltonian at a time-scale faster than the bath-correlation time result in non-Markovian anti-Zeno dynamics (AZD) \cite{kofman00acceleration, kofman01universal, kofman04unified, erez08, alvarez10zeno}, which allows the WM to exchange energy with a bath even out of resonance, thereby enhancing the heat currents significantly. Such periodic modulation has been realized experimentally \cite{almog11direct}, and previously been shown to enhance power in continuous thermal machines \cite{mukherjee20anti}. However, the application of AZD to enhance the performance of stroke thermal machines is still an unexplored subject. Here we realize an Otto cycle undergoing AZD; we show that the power in the AZD regime shows step-like behavior. AZD may enhance, as well as reduce the output power, with respect to that obtained in the Markovian dynamics limit. However, judicious choice of modulation time scales allow us to operate a thermal machine exhibiting significant quantum advantage, through generation of quantum enhanced power or refrigeration, without loss of efficiency or coefficient of performance, respectively.\\\\ 
The paper is organized as follows: in Sec. \ref{sec_qhm_gen} we discuss the dynamics of a fast-driven Otto cycle modelled by a generic WM. We focus on a minimal Otto cycle modelled by a two-level system in Sec. \ref{ottotls}, discuss 
the dynamics of the thermalizing strokes in Sec. \ref{sectherm}, analyze the Markov dynamics limit in Sec. \ref{secM}, the anti-Zeno dynamics in Sec. \ref{secAZD}, and quantum refrigeration in 
Sec. \ref{secRef}. Finally, we conclude in Sec. \ref{secCon}.

\section{A generic quantum-enhanced Otto cycle}
\label{sec_qhm_gen}

We consider an Otto cycle, modelled by a generic WM, and powered by a hot and a cold thermal bath with temperatures $T_{\rm h}$ and $T_{\rm c} < T_{\rm h}$ respectively. 
One can describe the setup through the Hamiltonian $H$: 
\ba
H &=& H_{\rm S} + H_{\rm Bh} + H_{\rm Bc} + H_{\rm SB} \non\\
H_{\rm SB} &=& \lambda_{\rm h}(t)S\otimes B_{\rm h} + \lambda_{\rm c}(t)S\otimes  B_{\rm c}.
\label{hamilgen}
\ea
Here $H_{\rm S}$, $H_{\rm Bh}$, $H_{\rm Bc}$ and $H_{\rm SB}$ denote the Hamiltonians describing the system (WM), hot bath, cold bath and interaction between the WM and the two thermal baths, respectively. 
The Hermitian operator $S$ causes transitions between the energy levels of the WM,  while   $B_{\rm h}$ and   $B_{\rm c}$ act on the hot and the cold bath, respectively; 
$\lambda_{j}(t)$, $j = \{\rm h,~c\}$, are time-dependent scalars, denoting the interaction strength between the WM and the hot (h) and cold (c) baths. For an Otto cycle in absence of control, $\lambda_{\rm h,c} = 0$
during the unitary strokes, while a non-zero $\lambda_{j}$ leads to thermalization of the WM with the $j$-th bath during a non-unitary stroke (see below). On a related note, a continuous 
thermal machine is in general accompanied by 
$\lambda_{\rm h,c}(t) \neq 0$ for all time $t$ \cite{klimovsky15thermodynamics, kosloff13quantum}.

\begin{figure}[h]
\begin{center}
\includegraphics[width=\columnwidth]{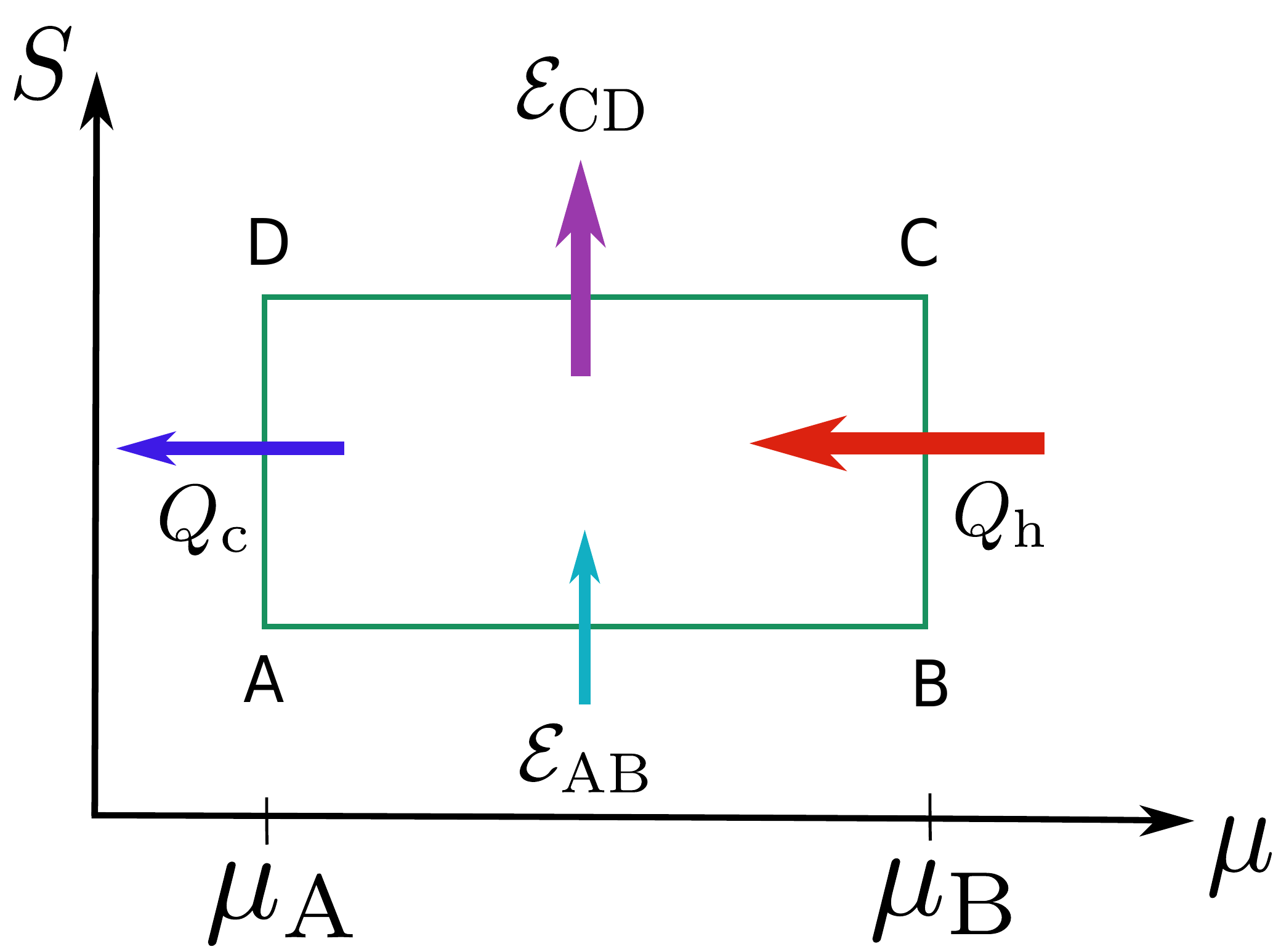}
\caption{Schematic diagram of an Otto cycle in the entropy $S$-$\mu$ plane. The cycle consists of two unitary strokes A to B and C to D, with energy flows $\mathcal{E}_{\rm AB}$ and $\mathcal{E}_{\rm CD}$, respectively, and two thermalization strokes  B to C and D to A, with  heat flows 
$\mathcal{Q}_{\rm h}$ and $\mathcal{Q}_C$, respectively.}
\label{otto-schematic}
\end{center}
\end{figure}

Below we describe one cycle of the Otto thermal machine considered here (see Fig. \ref{otto-schematic}) \cite{kosloff17the}.
\begin{itemize}
 \item {\bf First stroke:} We start with the WM in state $\rho_{{\rm S, A}}$, in equilibrium with the cold bath. The interaction strengths $\lambda_{\rm h,c}(t) = 0$ in this unitary stroke, such that 
 the WM is decoupled from both the baths. The system Hamiltonian $H_{\rm s}(\mu(t))$ is changed from $H_{\rm S}(\mu = \mu_A)$ at A to $H_{\rm S}(\mu = \mu_B)$ at B (see Fig. \ref{otto-schematic}) in a time interval $\tau_{\rm u1}$, where $\mu$
 is a time-dependent parameter 
 describing the Hamiltonian of the system. The state $\rho_{\rm S}(t)$ of the WM evolves in time following the von Neumann equation
 \ba
 \dot{\rho}_{\rm S}(t) = -i\left[H_{\rm S}(t), \rho_{\rm S}(t) \right].
 \label{eqvn}
 \ea
 Here for simplicity, unless otherwise stated, we consider $\hbar = k_{\rm B} = 1$.
 
 \item {\bf Second stroke:} In this non-unitary stroke of duration $\tau_{\rm h}$, the WM Hamiltonian is kept constant at $H_{\rm S} = H_{{\rm S}}(\mu_{\rm B})$ at B, $\lambda_{\rm c} = 0$, while the WM interacts with the hot bath through a non-zero $\lambda_{\rm h}(t)$. $\tau_{\rm h}$ is in general assumed to be large enough such that 
 the WM thermalizes with the hot bath at the end of this stroke at C. The dynamics of the WM during this stroke can be described by the master equation
 \ba
 \dot{\rho}_{\rm S}(t) = -i\left[H_{{\rm S}}(\mu_{\rm B}), \rho_{\rm S}(t) \right] + \mathcal{D}_{\rm h}(t)\left[\rho_{\rm S}(t) \right].
 \label{eqdisgen}
 \ea
  Here, $\mathcal{D}_{\rm h}(t)$ is a dissipative superoperator acting on the WM at time $t$ \cite{breuer02}. In general, for a WM evolving in presence of a thermal bath, and in absence of any time-dependent control Hamiltonian and constant $\lambda_{\rm h}$, 
$\mathcal{D}_{\rm h}$ is independent of time and describes a Markovian dynamics. However, as we show below, fast periodic control, in the form of rapid intermittent coupling / decoupling of the WM with the 
 hot bath, can lead to anti-Zeno non-Markovian dynamics, with time-dependent $\mathcal{L}_{\rm h}$ \cite{kofman00acceleration, kofman04unified, erez08, mukherjee20anti}.
 
 \item {\bf Third stroke:} Once again, we set $\lambda_{\rm h,c}(t) = 0$, while $H_{\rm S}(t)$ is changed from $H_{{\rm S}}(\mu_{\rm B})$ at C back  to $H_{{\rm S}}(\mu_{\rm A})$
at D, in a time interval $\tau_{\rm u2}$. The WM evolves following the von Neumann equation \eqref{eqvn} during this unitary stroke.
 
 \item {\bf Fourth stroke:} In this stroke of time duration $\tau_{\rm c}$, the WM Hamiltonian is kept constant at $H_{\rm S} = H_{{\rm S}}(\mu_{\rm A})$ at D, $\lambda_{\rm h} = 0$, while a 
 non-zero $\lambda_{\rm c}(t)$ allows the system to thermalize with the cold bath. Analogous to the second stroke, the WM evolves following Eq. \eqref{eqdisgen}, 
 with $\mu_{\rm B}$ and $\mathcal{D}_{\rm h}$ replaced by 
 $\mu_{\rm A}$ and $\mathcal{D}_{\rm c}$, respectively. At the end of this stroke,  the WM returns to its initial state  $\rho_{{\rm S, A}}$ at A, thereby completing the cycle.  
\end{itemize}
The cycle period is given by $\tau = \tau_{\rm u1} + \tau_{\rm h} + \tau_{\rm u2} + \tau_{\rm c}$.
We operate the thermal machine in the limit cycle, such that the WM reaches thermal equilibrium with the bath at the end of each non-unitary stroke.
The average energy $\langle E_{\alpha} \rangle = {\rm Tr}\left[\rho_{{\rm S},{\alpha} } H_{{\rm S},{\alpha} } \right]$ of the WM at the $\alpha$-th point (${\alpha}  = A, B, C, D$) allows us to obtain the heat $Q_{\rm h}$ and $Q_{\rm c}$, exchanged with 
the hot and the cold bath respectively, as,
\ba
Q_{\rm h} = \left(\langle E_{\rm C} \rangle - \langle E_{\rm B} \rangle\right)\non\\
Q_{\rm c} = \left(\langle E_{\rm A} \rangle - \langle E_{\rm D} \rangle\right),
\label{Qgen}
\ea
while the energy flows $\mathcal{E}_{\rm AB}$ and $\mathcal{E}_{\rm CD}$ during the first and third strokes are given by (Cf. Fig. \ref{otto-schematic})
\ba
\mathcal{E}_{\rm AB} = \left(\langle E_{\rm B} \rangle - \langle E_{\rm A} \rangle\right)\non\\
\mathcal{E}_{\rm CD} = \left(\langle E_{\rm D} \rangle - \langle E_{\rm C} \rangle\right),
\label{Qgen}
\ea
Energy conservation gives the total work $\mathcal{W}$ output, and the cycle-averaged power output as, 
\ba
\mathcal{P} = \frac{\mathcal{W}}{\tau} = \frac{\mathcal{E}_{\rm AB}+ \mathcal{E}_{\rm CD}}{\tau} = -\frac{Q_{\rm h} + Q_{\rm c}}{\tau},
\label{powgen}
\ea
and the efficiency as
\ba
\eta = -\frac{\mathcal{W}}{Q_{\rm h}}.
\label{effgen}
\ea
Here we have used the sign convention that energy flow (heat, work) is positive (negative) if it enters (leaves) the WM. A heat engine is characterized by $Q_{\rm h} > 0, Q_{\rm c} < 0, \mathcal{W} < 0$, while $Q_{\rm h} < 0, Q_{\rm c} > 0, \mathcal{W} > 0$ denotes the refrigerator regime, and we get the heat distributor regime for $Q_{\rm c} < 0, \mathcal{W} > 0$  \cite{mukherjee16speed, binder18book}.

\begin{figure}[h]
\begin{center}
\includegraphics[width = \columnwidth]{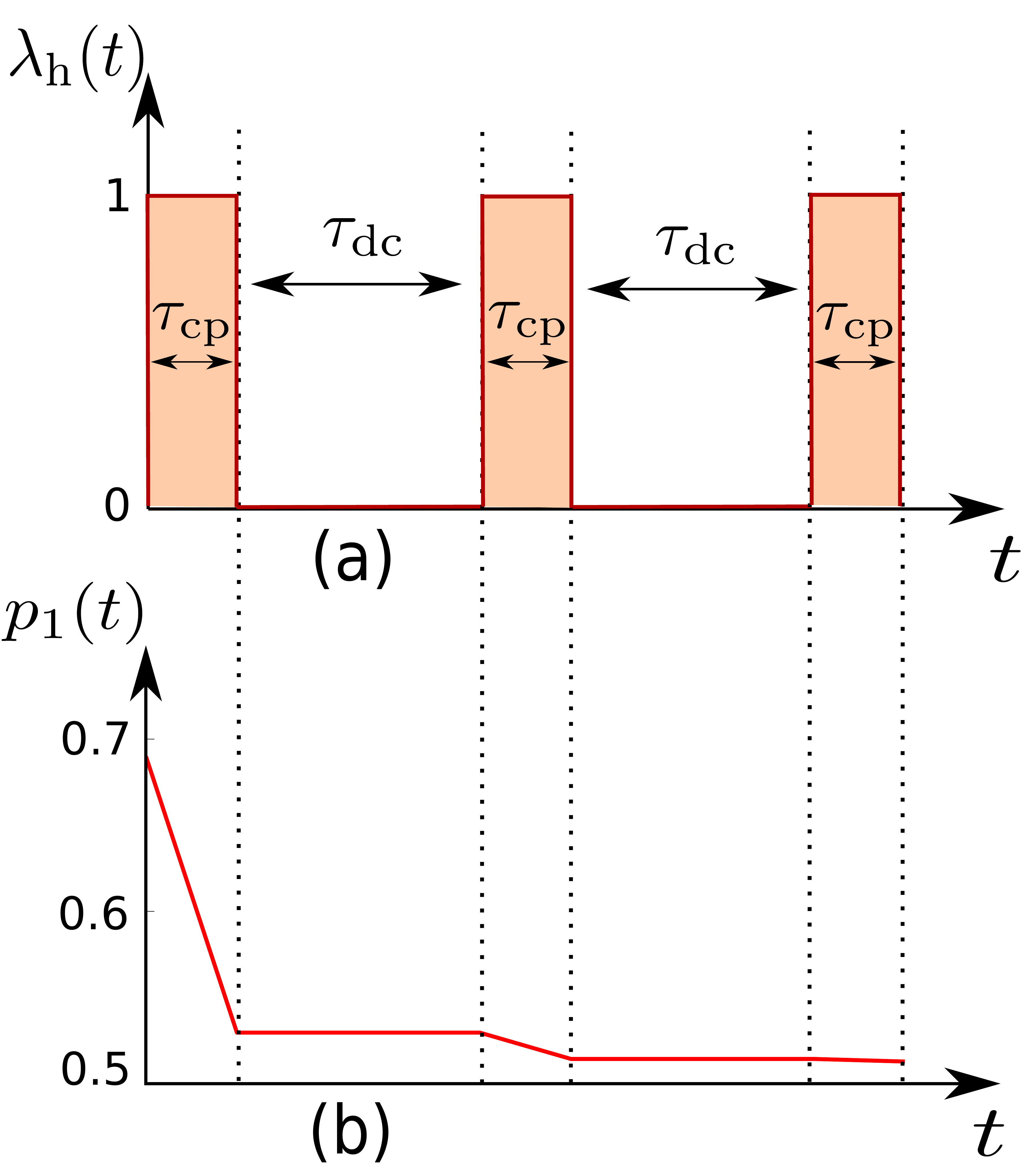}
\caption{(a) The interaction strength $\lambda_{\rm h}(t)$ between the WM and the hot bath and (b) the corresponding time-dependent occupation probability $p_1(t)$ for a two-level system WM (see Eq. \eqref{p1f}), during a thermalization stroke with the hot bath. $\lambda_{\rm c}(t)$ shows similar variation with time, during the thermalization stroke with the cold bath (not shown here).  Lorentzian bath spectrum has been used with $\delta=2$ and $\Gamma=0.4$. Initial state of the WM is the thermal state corresponding to the cold bath with $\beta_{\rm c}=0.01$ and $\omega_{\rm c}=80$. Here $\beta_{\rm h}=0.0005$, $\omega_{\rm h}=100$, $\tau_{\rm cp} = 1.5$ and $\tau_{\rm dc} = 4$.}
\label{step}
\end{center}
\end{figure}

As mentioned above, in general, for a setup subjected to time-independent Hamiltonian $H$ during the non-unitary strokes, one can use Born, Markov and secular approximations to arrive at
a time-independent dissipative Lindblad superoperator $\mathcal{L}_j$ ($j = \{\rm h,~c\}$) decribing the dynamics of the WM \cite{breuer02}. However, a $H(t)$ changing rapidly with time may invalidate
the Markov approximation, thereby leading to a time-dependent $\mathcal{L}_j(t)$, and a possibly non-Markovian dynamics \cite{breuer02, rivas12open, rivas14quantum}. Below, 
we harness this breakdown of Markovianity to achieve quantum 
advantage; we introduce a modification in the conventional Otto cycle \cite{kosloff17the}, in the form of fast periodic coupling, decoupling of the WM with the
thermal baths during the non-unitary strokes, implemented through step function forms of $\lambda_{j}(t)$. At the beginning of a non-unitary stroke, we couple the WM with a bath $j$ and
allow it to thermalize for a time interval $\tau_{\rm cp}$, during which time $\lambda_{j}(t)$ assumes a constant value 
$\bar\lambda_{j} > 0$. 
The coupling time interval is followed by decoupling of the WM and the bath, for a time interval $\tau_{\rm dc}$, realized through $\lambda_{j}(t) = 0$. 
Following the decoupling interval, we once more couple the WM with the bath for
a time interval $\tau_{\rm cp}$ ($\lambda_{j}(t) = \bar\lambda_{j}$), and repeat the above process till the WM thermalizes with the bath (see Fig. \ref{step}). 

One can show that rapid coupling / decoupling of the WM with a bath results in the WM evolving in time following the master equation (see Appendix \ref{gme}):
\ba
\dot{\rho}_{\rm S}(t) &=& \mathcal{D}_j\left[\rho_{\rm S}(t) \right] = \sum_{\omega}\tilde{\mathcal{R}}_j(\omega,t)  \mathcal{L}_{j, \omega}\left[\rho_{\rm S}(t) \right] + \text{h.c.};\non\\
\tilde{\mathcal{R}}_j(\omega,t) &\equiv& \int_{-\infty}^{\infty} d\nu G_j(\nu)\Big[\frac{\sin\left[\left(\nu - \omega \right)t\right]}{\nu - \omega} \non\\
&\pm& i \left(\frac{\cos\left[\left(\nu - \omega \right)t\right] - 1}{\nu - \omega}\right)\Big].
\label{rwaTsinc1}
\ea
Here the dissipative  superoperator $\mathcal{D}_j$ can be written in terms of its $\omega$-spectral 
components of Lindblad dissipators $\mathcal{L}_{j,\omega}$ (see below),  and $\tilde{\mathcal{R}}_j(\omega, t)$ (see Eq. \eqref{rwaTsinc1}). In case of $\rho_{\rm S}(t)$ that is diagonal in the energy basis, as can be expected for Otto cycles powered by thermal baths, and in presence of system Hamiltonians satisfying $\left[H_{\rm S}(t), H_{\rm S}(t^{\prime})\right] = 0$ for all times  $t, t^{\prime}$, one can show that
the dynamics is dictated by the coefficients 
$\mathcal{R}_j(\omega,t) \equiv {\rm Re}\left[\tilde{\mathcal{R}}_j(\omega,t)\right]$ \cite{shahmoon13engineering, mukherjee20anti}. The scalar $\mathcal{R}_j(\omega,t)$ is given by the convolution of the bath spectral response function $G_j(\nu)$, with spectral width $\sim \Gamma_{\rm B} \sim 1/\tau_{\rm B}$,
and the function $\frac{\sin\left[\left(\nu - \omega \right)t\right]}{\nu - \omega}=t\sinc((\nu-\omega)t)$, . Here we will consider the Kubo-Martin-Schwinger (KMS) condition \cite{breuer02}:
\ba
G_j(-\nu) = \exp\left[-\nu \beta_j \right]G_j(\nu).
\label{eqkms}
\ea 
As we discuss below, the dynamics of the thermal machine crucially depends on $\mathcal{R}_j(\omega,t)$, through the  time-energy uncertainty 
relation of quantum mechanics.

We show that choosing a $\tau_{\rm cp}  \lesssim \tau_{\rm B}$
may lead to the anti-Zeno dynamics, i.e., to a significant enhancement in the overlap between the sinc functions and the bath spectral functions, or equivalently, in the convolution
$\mathcal{R}_j(\omega,t)$. This in turn boosts the rate of heat flow between the WM and the $j$-th bath \cite{kofman00acceleration, erez08, mukherjee20anti}. On the other hand, the
effect of anti-Zeno boost in the rate of heat flow may be counteracted by the time intervals
$\tau_{\rm dc}$ during which the WM is kept decoupled 
from the thermal baths, and consequently associated with zero heat flow.
However, as we show below, judicious choice of parameters can allow us to engineer an Otto machine exhibiting significant quantum advantage, through a net reduction of thermalization 
time $\tau_{\rm th}$ for approximately the same amount of output work, 
and a resultant enhancement in output power (see Figs. \ref{heat_engine_a} and \ref{heat_engine_b}), or in refrigeration (see Figs. \ref{refri_a} and \ref{refri_b}).

On the other hand, long WM-baths coupling durations (i.e., $\tau_{\rm cp} \gg \tau_{\rm B}$) result in the ${\rm sinc}$ functions assuming the form of  delta functions. Consequently, we arrive at the standard Markovian form of the master equation \eqref{rwaTsinc1} describing the dynamics of conventional Otto thermal machines in absence of control, with time-independent $\mathcal{R}_j(\omega,t)$, given by $\mathcal{R}_j(\omega,t) = \pi G_j\left(\omega\right) > 0$.

\section{A fast-modulated minimal Otto cycle}

\subsection{Model}
\label{ottotls}

Here we exemplify the generic results discussed above, by focussing on the specific example of an Otto cycle involving a two-level system WM, described by the Hamiltonian 
\ba
H_{\rm S}(t)&=&\frac{\omega(t)}{2} \sigma_z, \non\\
H_{\rm Sh}&=& \lambda_{\rm h}(t)\sigma_x\otimes B_{h};~~H_{\rm Sc}= \lambda_{\rm c}(t)\sigma_x\otimes B_{c}.
\label{sysHam}
\ea 
Here $\sigma_{\alpha}$ denotes the Pauli matrix acting on 
the WM, along the 
$\alpha = x, y, z$ axis.

As detailed above for the general case, we  consider the WM to be prepared in the state $\rho_{{\rm S, D}}$, in thermal equilibrium with the cold bath, at the start of the first stroke of a cycle. 
The frequency $\omega(t) > 0$ is modulated from $\omega_{\rm c}$ to $\omega_{\rm h} > \omega_{\rm c}$, while $\lambda_{\rm h,c} = 0$ during the first stroke, during which time the state of the WM remains unchanged, so that $\rho_{{\rm S,B}}=\rho_{{\rm S, A}}$, as can be seen from Eqs. \eqref{eqvn} and \eqref{sysHam}. The WM is allowed to
thermalize with the hot bath at constant $\omega(t) = \omega_{\rm h}$ and $\lambda_{\rm c} = 0$, during the second non-unitary stroke. We consider a step-function $\lambda_{\rm h}(t)$ during this stroke, as shown in Fig. \ref{step}a. 
The frequency is again reduced to from $\omega_{\rm h}$ to $\omega_{\rm c}$ during the third unitary stroke, during which time the state of the WM remains unchanged. Finally, the WM is allowed to thermalize with the cold bath following  a step-function $\lambda_{\rm c}$ and $\lambda_{\rm h} = 0$ during the fourth thermalization stroke, such that the cycle is completed. For simplicity, here we take $\bar{\lambda}_{\rm h,c}$ to be unity.

\subsection{Thermalization strokes}
\label{sectherm}

We now analyze the dynamics of the WM during a non-unitary stroke, in presence of a step-function $\lambda_j(t)$, as shown in Fig. \ref{step}. 
One can use the time-dependent occupation probabilities $p_1(t)$ and $p_2(t)$, of the states $\ket{0}\bra{0}$ and $\ket{1}\bra{1}$, respectively, to write (see Appendix \ref{appB})
\begin{eqnarray}
\nonumber
\rho_{\rm S}(t) &=& p_1(t)\ket{0}\bra{0}+p_2(t)\ket{1}\bra{1} \non\\
\dot{p_1}(t) &=& 2\lambda_j(t)^2\left[\mathcal{R}_{j}(\omega_{j}, t)p_2(t)-\mathcal{R}_{j}(-\omega_{j}, t)p_1(t)\right]\non\\
\dot{p_2}(t) &=& - \dot{p_1}(t).
\label{rate}
\end{eqnarray}
A $\mathcal{R}_j(\pm \omega_j, t) > 0$ for all times $t$ signify Markovian dynamics. On the other hand, non-Markovian dynamics ensues for $\mathcal{R}_j(\pm \omega_j, t)$ assuming negatives values for some time-intervals (see Fig. \ref{RjNM}) \cite{chruscinski10non, rivas14quantum}.

During the coupling time-intervals ($\lambda_j(t) = \bar{\lambda}_j=1$), the above rate equations  \eqref{rate} result in the occupation probabilities
\begin{eqnarray}
\label{p1f}
p_1(t) &=& \frac{e^{-(J_{j}^{+}+J_{j}^{-})}[J_{j}^{-} \bar{p}_1-J_{j}^{+} \bar{p}_2]+J_{j}^{+}}{(J_{j}^{+}+J_{j}^{-})}\non \\
p_2(t) &=& \frac{e^{-(J_{j}^{+}+J_{j}^{-})}[-J_{j}^{-} \bar{p}_1+J_{j}^{+} \bar{p}_2]+J_{j}^{-}}{(J_{j}^{+}+J_{j}^{-})},
\end{eqnarray}
where $\rho_{\rm S}(t_0) = \bar{p}_1\ket{0}\bra{0}+\bar{p}_2\ket{1}\bra{1}$ corresponds to the initial state at the beginning of a coupling time-interval $\tau_{\rm cp}$, when the WM starts interacting with  the $j$-th bath.
Here, 
\ba
J_{j}^{\pm}(t_0,t) &=& 2\int_{t_0}^t \mathcal{R}_{j}(\pm \omega_{j}, t')dt' \non\\
\mathcal{R}_{j}(\pm \omega_{j}, t) &=& {\rm Re}\left[\tilde{\mathcal{R}}_{j}(\pm \omega_{j}, t)\right] \non\\ &=&  \int_{-\infty}^{+\infty}d\nu G_j(\nu)\frac{\sin(\nu \mp \omega_{j})t}{\nu \mp \omega_{j}}\non\\
&=&\int_{-\infty}^{+\infty}d\nu G_j(\nu) t\sinc((\nu\mp\omega_j)t).
\label{eqsinc}
\ea
As seen from Eq. \eqref{rate}, the condition 
\ba
\frac{p_{1}(t)}{p_{2}(t)}=\frac{\mathcal{R}_{j}(\omega_{j}, t)}{\mathcal{R}_{j}(-\omega_{j}, t)}
\label{statess}
\ea
leads to the steady state with $\dot{p}_1(t) = \dot{p}_2(t) = 0$, for the $j$-th bath.
The general expressions for heat (Eq. \eqref{Qgen}) get reduced to, 
\ba
\label{expreheat}
Q_{\rm h} =\omega_{\rm h}(p_{1,1}-p_{1,2})\non\\
Q_{\rm c} =\omega_{\rm c}(p_{1,3}-p_{1,4}),
\ea
where $p_{1,\alpha}$ denotes the occupation probability of the state $\ket{0}\bra{0}$, after the end of the stroke $\alpha$ of a cycle.

We note that Eqs. \eqref{p1f} - \eqref{statess} describe the dynamics of the WM only during the time-intervals $\tau_{\rm cp}$, when the WM is coupled to a bath. In contrast, during the decoupling time-intervals $\tau_{\rm dc}$ with $\lambda_j(t) = 0$, $\rho_{\rm S}$ does not evolve with time, and we have $\dot{p_1}(t) = \dot{p_2}(t) = 0$ (see Eq. \eqref{rate} and Fig. \ref{step}).

We note that in general a system coupled to a thermal bath equilibrates with the bath asymptotically, reaching the corresponding exact thermal (Gibbs) state only at infinite time \cite{breuer02}. Therefore,
in order to realize a practical thermal machine, we consider the WM to be thermalized with a bath $j$ at temperature $T_j$ ($=1/\beta_j$), as long 
as it is within a small $\epsilon$ distance from the thermal (Gibbs) state $\rho_{{\rm th},j} = \exp\left[-\beta_j H_{\rm S,\alpha}\right]/Z_j$, $Z_j$ being the corresponding partition function \cite{mukherjee13speeding}. Here we quantify the distance  between two states $\rho = p_1\ket{0}\bra{0} + p_2\ket{1}\bra{1}$ and $\rho^{\prime} = p_1^{\prime}\ket{0}\bra{0} + p_2^{\prime}\ket{1}\bra{1}$ as $\epsilon = |p_1 - p_1^{\prime}| = |p_2 - p_2^{\prime}|$. 
\subsection{Markov Limit}
\label{secM}
The dynamics of the WM depends on the interplay between the bath correlation time $\tau_{\rm B}$, the thermalization time and the coupling time interval $\tau_{\rm cp}$.
 Markov approximation is valid in the limit $\tau_{\rm cp} \gg \tau_{\rm B}$, when the sinc functions inside the 
integrals in Eq. \eqref{eqsinc} reduce to delta functions, leading to $\mathcal{R}_{j}(\pm \omega_{j}) = \pi G_{j}(\pm \omega_{j})$. 
Consequently, the heat flows and the power (see Eqs. \eqref{Qgen} and \eqref{powgen}) assume finite values only for finite $G_{j}(\omega_{j})$, i.e., for thermal baths which are at resonance with the WM.
On the other hand, for a generic thermal bath sufficiently detuned from WM, such that $G_{j}(\omega_{j}) \approx 0$, we get $\dot{p_1}(t), \dot{p_2}(t) \approx 0$ (see Eqs. \eqref{rate} and \eqref{eqsinc}), and consequently vanishingly small heat flows $Q_{\rm h},~Q_{\rm c}$ (Cf. Eq. \eqref{expreheat}), and  the output power $\mathcal{P} = -\left(Q_{\rm h} + Q_{\rm c} \right)$ (see Figs. \ref{heat_engine_a} and \ref{heat_engine_b}).

On a related note, the KMS condition \eqref{eqkms} determines the steady state Eq. \eqref{statess}, given by 
\ba
\frac{p_{1}(t)}{p_{2}(t)}=\exp\left[\omega_{j} \beta_j \right].
\label{statessM}
\ea

\subsection{Anti-Zeno limit}
\label{secAZD}
We now focus on the regime $\tau_{\rm cp} \lesssim \tau_{\rm B}$, such that timescales shorter than the bath-correlation time become relevant. In this limit, the sinc functions in Eq. \eqref{eqsinc} cease to be 
delta functions anymore; instead, they assume finite widths $\Delta \nu \sim 1/t$ centered around $\nu = \omega_{j}$, thus giving rise to time-dependent $\mathcal{R}_j(\omega_{j}, t)$ and 
$\mathcal{R}_j(-\omega_{j}, t)$ (see Fig. \eqref{RjNM}).
This broadening of the sinc functions is a direct consequence of time-energy uncertainty relation of quantum mechanics, arising due to small $\tau_{\rm cp}$. Incredibly, this 
fast coupling / decoupling of the WM and the baths lead to AZD, such that the WM may thermalize with the $j$-th bath at a finite rate, even for the corresponding bath spectral function (see Figs. \ref{Fig6whole}- \ref{therm_ohmic}) peaking at a frequency $\omega_{j} + \delta \neq \omega_{j}$, and $G_j(\omega_{j}) \approx 0$, due to significant enhancement 
in values of 
the integrals in Eq. \eqref{eqsinc}. The finite thermalization times in turn boost the cycle-averaged heat currents, power and refrigeration, as compared to the Markovian limit of $\tau_{\rm cp} \gg \tau_{\rm B}$.

One may engineer AZD by implementing the following protocol during the thermalization strokes: the WM is to be coupled with the thermal bath for a time interval $\tau_{\rm cp} \lesssim \tau_{\rm B}$. 
Following this coupling period, the WM is decoupled from the bath for a time-interval $\tau_{\rm dc} \gg \tau_{\rm B}$, such that 
all system-bath correlations are destroyed. The WM is then recoupled with the bath, and the above process is  repeated, till the WM reaches the desired thermal state. 

We note that a fair comparison between the Markovian and the AZD regime demands the corresponding steady states (see Eq. \eqref{statess}) to be approximately same. This is indeed the case for 
\ba
\beta_{\rm h,c} \ll \tau_{\rm cp}~~~\text{and}~~~\tau_{\rm cp}^{-1} < \omega_{j},
\label{eqcond}
\ea
such that, 
\begin{equation}
\label{criteria}
\frac{\mathcal{R}_j(-\omega_{j}, t)}{\mathcal{R}_j(\omega_{j}, t)}  \approx e^{-\beta_{j}\omega_{j}} = \frac{G_{j}(-\omega_{j})}{G_{j}(\omega_{j})}.
\end{equation}
\begin{figure}[h]
\begin{center}
\subfloat[]{\includegraphics[height = 0.65\columnwidth, width = \columnwidth]{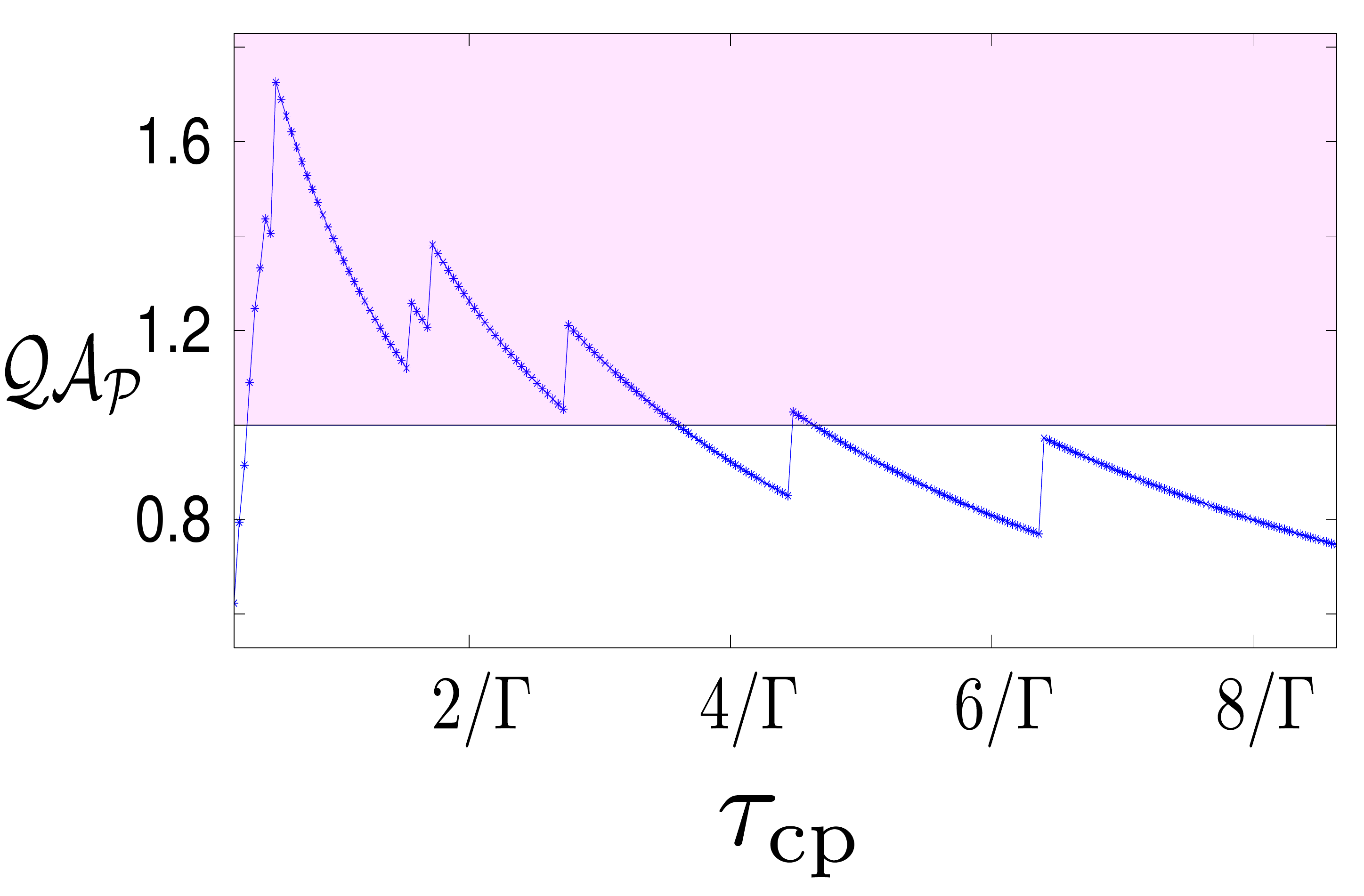}
\label{heat_engine_a}}
\hfill
\subfloat[]{\includegraphics[height = 0.65\columnwidth, width = \columnwidth]{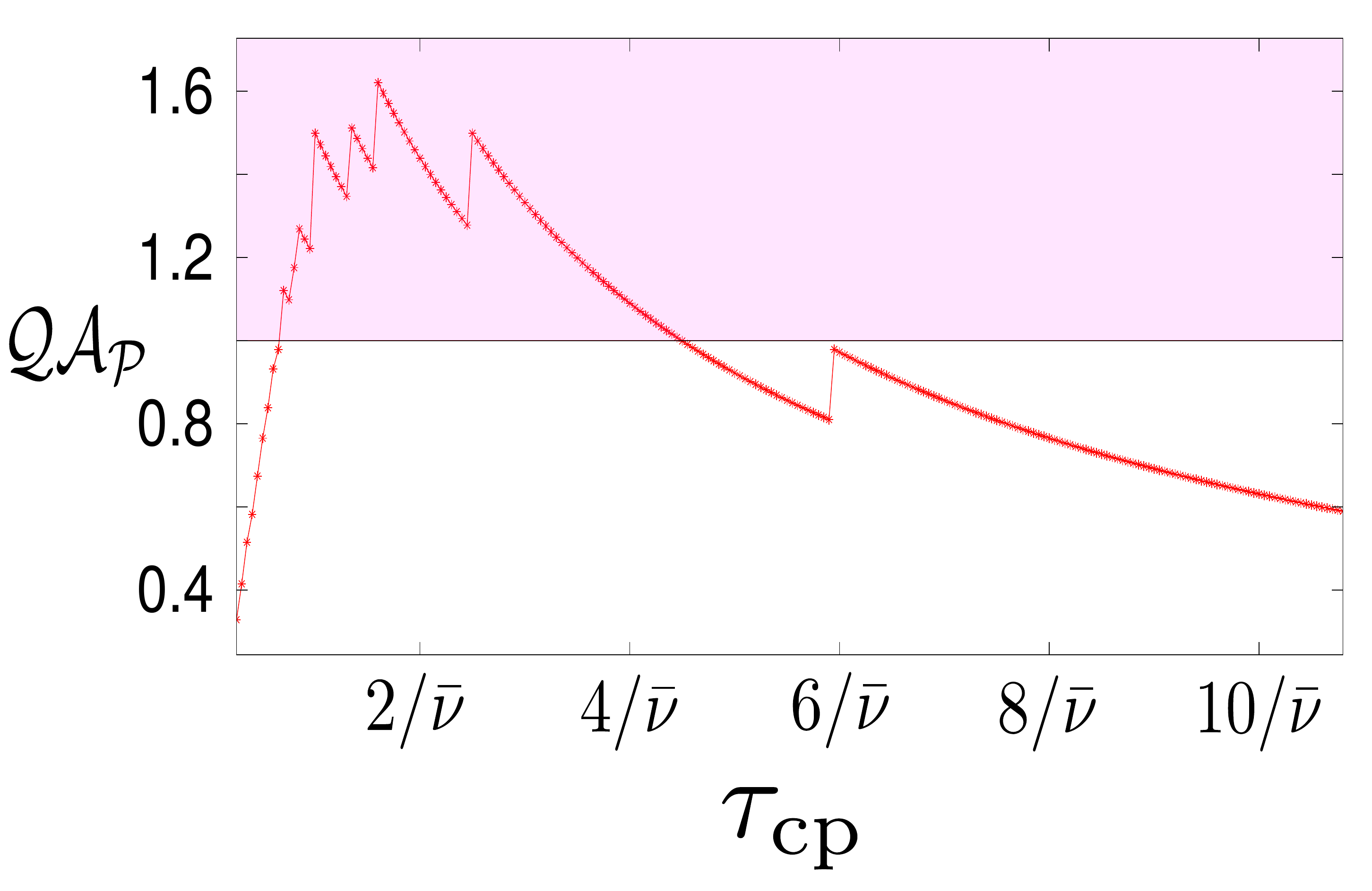}
\label{heat_engine_b}}
\caption{Quantum advantage ratio for the cycle-averaged output power of heat engine (see Eq.\eqref{QApower}) with (a) Lorentzian bath spectral function (see Eq. \eqref{bath1})  for $\delta=2$, $\Gamma=0.4$ and $\gamma_0 = 1$, and with (b) super-Ohmic bath spectral function (see Eq. \eqref{bath2}) for $\delta=0.1$, $\bar{\nu}=0.5$ and $\gamma_0 = 1$. We get quantum advantage for $\mathcal{QA}_{\mathcal{P}} > 1$, shown here by the shaded region. Here  $\beta_{\rm h} =0.0005$, $\beta_{\rm c}=0.01$, $\omega_{\rm c}=80$, $\omega_{\rm h}=100$, $\epsilon < 0.0015$,  $\tau_{\rm u1} = \tau_{\rm u3}= 0.4/\Gamma=0.5/\bar{\nu}$, $\tau_{\rm dc}= 1.6/\Gamma=2/\bar{\nu}$. Horizontal line indicates $\mathcal{QA}_{\mathcal{P}} = 1$.}
\label{Fig3}
\end{center}
\end{figure}

We compare the cycle-averaged power $\mathcal{P}(\tau_{\rm cp})$ (see Eq. \eqref{powgen}) for $\tau_{\rm cp} < \tau_{\rm th}$ and $\mathcal{P}_{\rm M}$ in the Markovian regime, for heat engines operated in presence of thermal baths with Lorentzian (cf. Fig. \ref{heat_engine_a}) and super-Ohmic (cf. Fig. \ref{heat_engine_b}) bath spectral functions (see Apps. \ref{appC} and \ref{appD}). To this end, we define the quantum advantage ratio 
\ba
\mathcal{QA}_{\mathcal{P}} = \frac{\mathcal{P}(\tau_{\rm cp})}{\mathcal{P}_{\rm M}}.
\label{QApower}
\ea
A $\mathcal{QA}_{\mathcal{P}} > 1$ indicates a quantum advantage through AZD induced enhancement of cycle-averaged output power, as compared to the Markovian limit. On the other hand, $\mathcal{QA}_{\mathcal{P}} < 1$ implies the time-energy uncertainty relation during the AZD fails to yield any quantum advantage. One can understand the behavior of $\mathcal{QA}_{\mathcal{P}}$ in Figs. \ref{heat_engine_a} and \ref{heat_engine_b}  by noting that
 small $\tau_{\rm cp}$ enhances the rate of heat flow between the WM and a thermal bath, through broadening of the corresponding sinc function. On the other hand,  every $\tau_{\rm cp}$ is followed by a decoupling time interval $\tau_{\rm dc}$, till the WM thermalizes with the bath, during which times heat flow ceases between the WM and the bath. Consequently, the power, which is function of $\tau_{\rm cp}$, $\tau_{\rm dc}$ and the total number of coupling and decoupling time-intervals, do not vary monotonically with decreasing $\tau_{\rm cp}$. Rather, the duration $\tau_{dc}$ of each decoupling time-interval and the total number $\mathcal{N}_{\rm dc}$ of decoupling time intervals remaining 
 constant, power increases initially as 
$\tau_{\rm cp}$ is decreased, owing to the enhancement in heat flow during the coupling time-intervals. However, smaller $\tau_{\rm cp}$, at a constant $\tau_{\rm dc}$, may demand higher number of coupling / decoupling time-intervals in order for the system to thermalize. Consequently,
 the power increases with decreasing $\tau_{\rm cp}$ as long as $\mathcal{N}_{\rm dc}$ (and hence the total decoupling time duration  $\mathcal{N}_{\rm dc}\tau_{dc}$) remain constant, while they may show sharp drops for increasing $\mathcal{N}_{\rm dc}$.  However, as seen from Figs. \ref{heat_engine_a} and \ref{heat_engine_b}, one can achieve significant quantum advantage through proper choice of small $\tau_{\rm cp}$. 
 
 The exact values of $\tau_{\rm cp}$ where $\mathcal{QA}_{\mathcal{P}}$ show spikes depend non-trivially on the setup and control parameters, through Eq. \eqref{powgen} and Eqs. \eqref{rate} - \eqref{expreheat}. However, as one can see from Fig. \ref{RjNM}, $\mathcal{R}_j(\omega_j,t)$ varies weakly with time at large $t$. Consequently the thermalization times (see Figs. \ref{therm_lorentz} and \ref{therm_ohmic}), and hence $\mathcal{QA}_{\mathcal{P}}$ (Figs. \ref{heat_engine_a} and \ref{heat_engine_b}), show smoother variations with $\tau_{\rm cp}$ at larger $\tau_{\rm cp}$, assuming spikes at approximately regular intervals, which scale as $\gamma^{-1}$. On the other hand, the strong time-dependence of $\mathcal{R}_j(\omega_j,t)$ for small $t$ translates to more  irregular behavior of $\mathcal{QA}_{\mathcal{P}}$ at shorter $\tau_{\rm cp}$, albeit with larger values of the quantum advantage ratios.

For the parameter values chosen in Figs. \ref{heat_engine_a} (Lorentzian
bath spectral function) $\mathcal{N}_{\rm dc}$ assumes a maximum value of
$14$ for the minimum duration of $\tau_{\rm cp}$ considered here ($\tau_{\rm cp}
= 0.2/\Gamma$), while the same for Fig. \ref{heat_engine_b} (super-Ohmic
bath spectral function) is $\mathcal{N}_{\rm dc} = 26$ for $\tau_{\rm cp} =
0.25/\bar{\nu}$. On the other hand, $\mathcal{N}_{\rm dc}$ reduces to zero in the Markovian limit of
$\tau_{\rm cp}$ of the order of the thermalization time, such that the WM
is always coupled with the corresponding bath during the thermalization
strokes.

The efficiency $\eta = 1 - \omega_{\rm c}/\omega_{\rm h}$, is 
independent of the details of 
the strokes, and rather depends only on the steady 
states. Consequently, the efficiencies are approximately identical for heat engines operating in the Markovian and the AZD regimes, as long as the conditions \eqref{eqcond} are satisfied. As a result,
the control protocol presented here allows us to realize a heat engine which delivers quantum enhanced power, without any loss of efficiency.

It is worthwhile to mention that in contrast to AZD, Zeno dynamics ensues for very small $\tau_{\rm cp}$ ($\tau_{\rm cp} \ll \tau_{\rm B}$), when the excessive broadening of the sinc functions results in decrease of power  with decreasing $\tau_{\rm cp}$ \cite{misra77the,itano90quantum, kofman00acceleration, mukherjee20anti}.

\subsection{Quantum Otto Refrigerator}
\label{secRef}

\begin{figure}[h]
\begin{center}
\subfloat[]{\includegraphics[height = 0.65\columnwidth, width = \columnwidth]{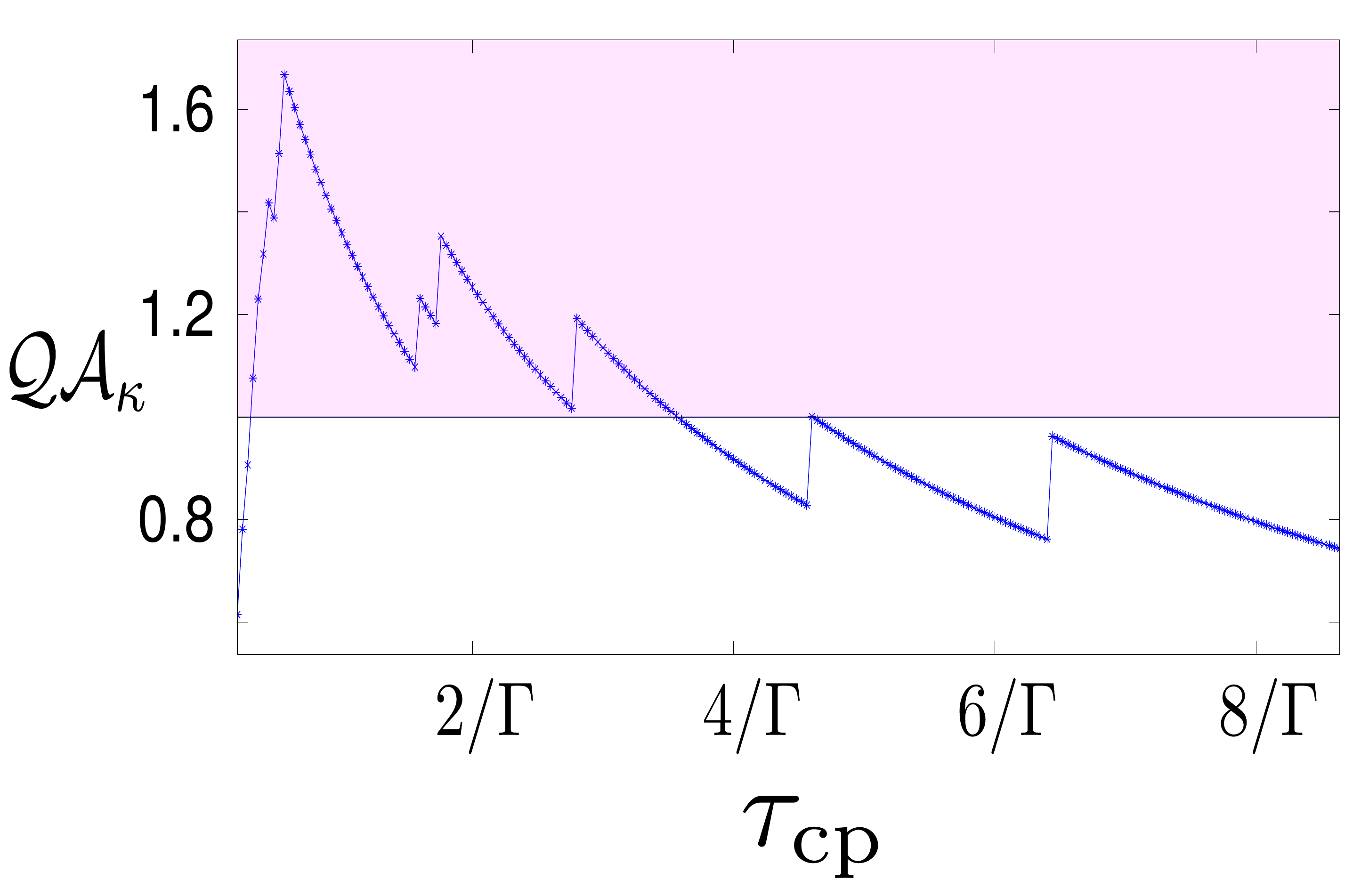}\label{refri_a}}
\hfill
\subfloat[]{\includegraphics[height = 0.65\columnwidth, width = \columnwidth]{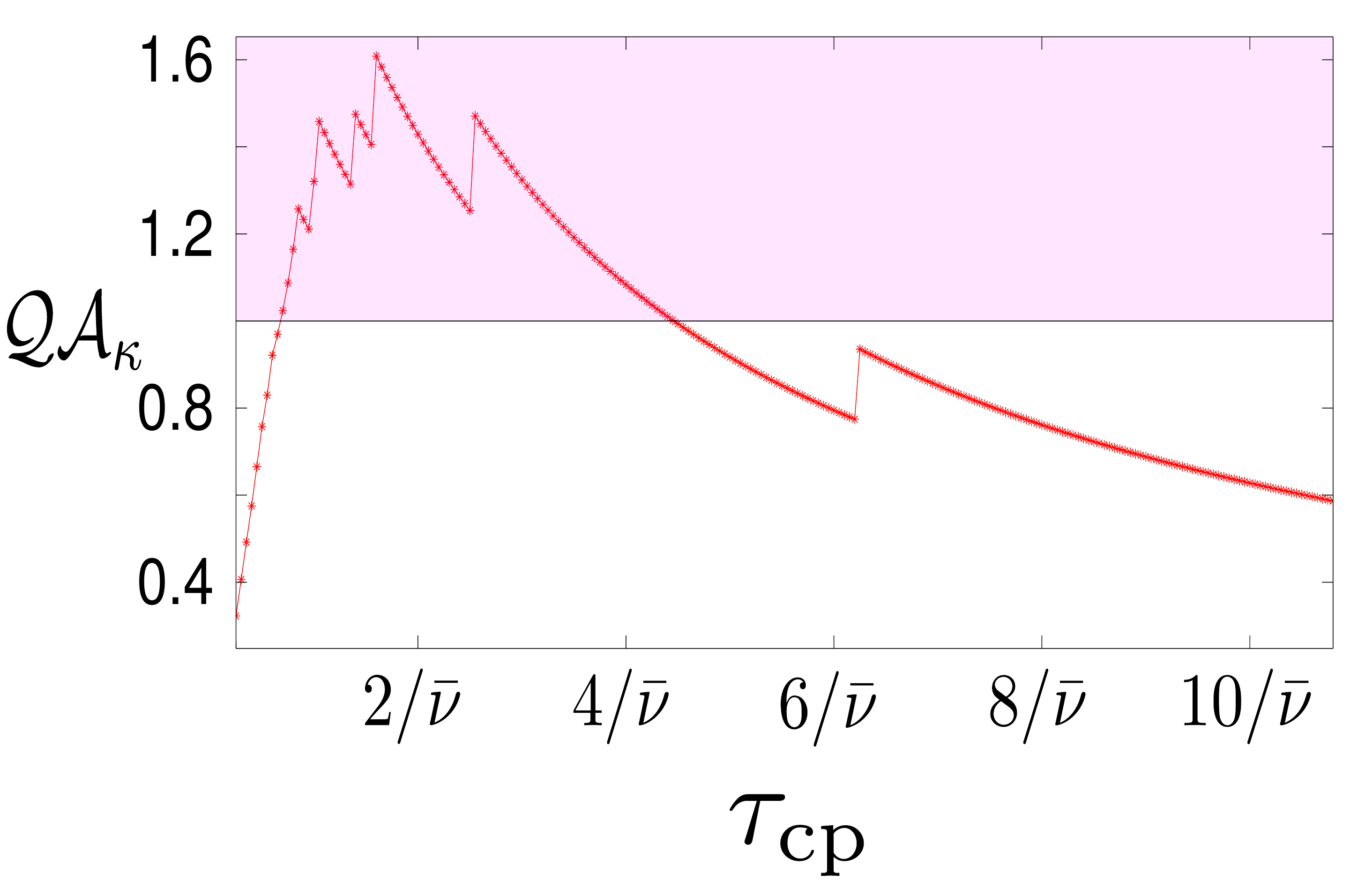}\label{refri_b}}
\caption{Quantum advantage ratio for the cycle-averaged cooling rate (see Eq. \eqref{QAkappa}) of a refrigerator with (a) Lorentzian bath spectral function (see Eq. \eqref{bath1}), for $\delta=2$, $\Gamma=0.4$ and $\gamma_0 = 1$,  and with (b) super-Ohmic bath spectral function (see Eq. \eqref{bath2}) for $\delta=0.1$,  $\bar{\nu}=0.5$ and $\gamma_0 = 1$. We get quantum advantage for $\mathcal{QA}_{\kappa} > 1$, shown here by the shaded region. Here $\beta_{\rm h} = 0.007$, $\beta_{\rm c} = 0.01$, $\omega_{\rm c} = 10$, $\omega_{\rm h} = 120$, $\epsilon < 0.0030$, $\tau_{\rm u1}=\tau_{\rm u3}= 0.4/\Gamma=0.5/\bar{\nu}$ and $\tau_{\rm dc}= 1.6/\Gamma=2/\bar{\nu}$, $\beta_{\rm h} = 0.007$. Horizontal line indicates $\mathcal{QA}_{\kappa} = 1$.}
\label{Fig4}
\end{center}
\end{figure}
One can operate the Otto cycle in the refrigerator regime as well, by choosing  \cite{kosloff14quantum, klimovsky13minimal, bhattacharjee20quantum} 
\ba 
\frac{\omega_{\rm h}}{\omega_{\rm c}}>\frac{T_{\rm h}}{T_{\rm c}}.
\label{condref}
\ea
The operation can be quantified through the cycle-averaged cooling rate $\kappa$:
\ba
\kappa = \frac{Q_{\rm c}}{\tau},
\ea
and the coefficient of performance CoP:
\ba
{\rm CoP} = \frac{Q_{\rm c}}{\left(\mathcal{E}_{\rm AB} + \mathcal{E}_{\rm CD}\right)}.
\ea 
As seen in the heat engine regime, a quantum refrigerator oprating with AZD ensues for \eqref{condref} and $\lambda_{\rm h,c}(t)$ of the form shown in Fig. \eqref{step}.
Consequently, one can achieve quantum advantage in the form of enhanced
$\kappa$ in the limit $\tau_{\rm cp} \lesssim \tau_{\rm B}$, 
at approximately the same CoP, as compared to an equivalent traditional Markovian Otto refrigerator, as long as Eq. \eqref{eqcond} is satisfied. Analogous to the heat engine regime, one can quantify the quantum advantage $\mathcal{QA}_{\kappa}$ through the ratio
\ba
\mathcal{QA}_{\kappa} = \frac{\kappa(\tau_{\rm cp})}{\kappa_{\rm M}},
\label{QAkappa}
\ea
where $\kappa(\tau_{\rm cp})$ and $\kappa_{\rm M}$ denote the cooling rates for $\tau_{\rm cp}<\tau_{\rm th}$ and the Markovian regime, respectively. As before, $\mathcal{QA}_{\kappa} > 1$ implies quantum advantage arising due to the time-energy relation of quantum mechanics (see Figs. \ref{refri_a} and Fig. \ref{refri_b}). 
In case of the refrigerator, we get a maximum $\mathcal{N}_{\rm dc} = 13$ for the minimum $\tau_{\rm cp} = 0.2/\Gamma$ considered in Fig. \ref{refri_a} (Lorentzian bath spectral function), while  $\mathcal{N}_{\rm dc}$ assumes a maximum value of $27$ for a minimum  minimum $\tau_{\rm cp} = 0.25/\bar{\nu}$ considered in Fig. \ref{refri_b} (super-Ohmic bath spectral function).

\section{Conclusion}
\label{secCon}
We have studied anti-Zeno dynamics in fast driven quantum otto cycles. We have shown how repeated decoupling and coupling of the WM and the thermal baths  during the non-unitary strokes 
can lead to non-Markovian anti-Zeno dynamics with significant enhancement in output power, in case of a heat engine, and cooling rate, in case of a refrigerator.  Yet, this quantum advantage, quantified by the ratios $\mathcal{QA}_{\mathcal{P}}$ (see Eq. \eqref{QApower} and Figs. \ref{heat_engine_a} and \ref{heat_engine_b}) and $\mathcal{QA}_{\kappa}$ (see Eq. \eqref{QAkappa} and Figs. \ref{refri_a} and Fig. \ref{refri_b}),  is 
non-monotonic with increasing frequency of modulation.  The energy flow between a bath and the WM is enhanced during the short coupling periods. On the other hand, the  decoupling time intervals are associated with zero heat flow. However, through proper choice of parameters, one can operate the 
cycle such that the AZD leads to an overall enhancement in the cycle-averaged power or cooling rate at the same efficiency or coefficient of performance, respectively. We emphasize that this improvement in performance is inherently quantum in nature;
the small time scale, obtained in the form of fast modulation during the non-unitary strokes, translates to increased  energy flow between the WM and the bath, even when they are not in resonance, owing to the time-energy uncertainty relation of quantum mechanics. 

We note that the control protocol presented above can be expected to significantly enhance the performance of a thermal machine only if the working medium is sufficiently detuned from the baths. On the other hand, for the special case of the working medium being at resonance with the baths, in general the heat currents are large even in absence of any control. Furthermore, under such a resonant condition, fast periodic coupling / decoupling of the WM and the baths can lead to Zeno effect, with subsequent reduction in output power or refrigeration \cite{misra77the, itano90quantum}.

It is also worthwhile to mention that as discussed in Sec. \ref{secAZD}, in order to have a fair comparision between the AZD limit and traditional Otto cycles operating in the Markovian limit,  here we have allowed the WM to thermalize with the bath at the end of a non-unitary stroke. However, one can also operate the machine without imposing this condition of WM-bath thermalization. For example, one can terminate the non-unitary stroke at the end of the first coupling time interval, such that the duration of the non-unitary stroke is  $\tau_{\rm cp}$. Such a protocol would reduce the loss incurred during the decoupling times, which might in turn enhance the output power (refrigeration rate) even further \cite{erdman19maximum}, at the cost of low output work (refrigeration) per cycle of the heat engine (refrigerator). However, a detailed analysis of such an operation protocol is beyond the scope of the current paper.

One can envisage experimental realizations through working mediums modelled by nano-mechanical oscillators \cite{klaers17squeezed}, single atoms \cite{rossnage16a}, or NV centers in diamonds \cite{klatzow19experimental}. The rapid coupling / decoupling of the WM and a thermal bath during the non-unitary strokes can be implemented by suddenly changing the energy level spacing of the WM, such that it becomes highly non-resonant with the thermal bath, thereby effectively stopping any energy flow between the two. Thereafter one can  again revert back the energy-level spacing to its initial value, thus effectively recoupling the WM with the thermal bath. 

We expect the control protocol presented here to find applications in modelling of quantum thermal machines exhibiting significant quantum advantage, and also to lead to further studies of similar control schemes in many-body quantum thermal machines \cite{campisi16the, chen19an, hartmann19many, revathy20universal}, and related technologies based on open quantum systems.

\appendix
\section{General master equation}
\label{gme}
We start with the time convolution-less master equation in the interaction picture,
\begin{align}
\nonumber
&& \dot{\rho}_{\rm S}(t)=-\lambda_j(t)^2\int_0^t Tr_{B_j}[S(t)\otimes B_j(t),\\
&& [S(s)\otimes B_j(s),\rho_{\rm S}(t)\otimes \rho_{B_j}]],
\label{eqtcl}
\end{align} 
where $S(t)\otimes B_j(t)=e^{iH_{\rm S}t}Se^{-iH_{\rm S}t}\otimes e^{iH_{\rm Bj}t}B_je^{-iH_{\rm Bj}t}$ ($j=\rm h,c$), with $H_{\rm SB}=\lambda_j(t)S\otimes B_j$, $S$ and $B_j$ being the system and bath operators respectively.
Expanding Eq. \eqref{eqtcl}, we get,
\begin{eqnarray}
\nonumber
&& \dot{\rho}_{\rm S}(t)={\lambda_j(t)}^2\Big[-\int_0^t ds(S(t)S(s)\rho_{\rm S}(t)\Phi(t-s))\\
\nonumber
&&+ \int_0^t ds(S(s)\rho_{\rm S}(t)S(t)\Phi(t-s))+\int_0^t ds(S(t)\rho_{\rm S}(t)S(s)\\
\label{master_first1}
&&\Phi(s-t))-\int_0^t ds(\rho_{\rm S}(t)S(s)S(t)\Phi(s-t))\Big],
\end{eqnarray}
where $\Phi(t-s)=Tr(\rho_{\rm B_j} B_j(t) B_j(s))$ is the bath correlation function, and 
\begin{eqnarray}
\nonumber
S(t)=S^{\dagger}(t);~~~ B_j(t)={B_j}^{\dagger}(t).
\end{eqnarray}
Additionally, replacing $(t-s)$ by $\tau$, one can write the first term inside the square bracket of the Eq. (\ref{master_first1}) as,
\begin{align}
-\sum_\omega S^{\dagger}(\omega)S(\omega) \rho_{\rm S}(t) \int_0^t e^{-i(\nu-\omega)\tau}d\tau\int_{-\infty}^{\infty} G_j(\nu) d\nu.
\label{eqT1}
\end{align}
where,
\ba
\Phi(t-s) &=& \Phi(\tau)=\int_{-\infty}^{+\infty}G_j(\nu)e^{-i\nu\tau} d\nu ~~~ \text{and} \non\\
S(t) &=& \sum_\omega S(\omega) e^{-i\omega t}.
\label{eqphi}
\ea
We have also used the Rotating Wave Approximation (RWA) \cite{breuer02} and the Hermiticity property $S(t)=S^{\dagger}(t)$, implying $\sum_\omega S(\omega)e^{-i\omega t}=\sum_\omega S^{\dagger}(\omega)e^{i\omega t}$.
Similarly, the second term inside the square bracket of Eq. (\ref{master_first1}) is,
\begin{equation}
\label{second_term}
\sum_\omega S(\omega) \rho_{\rm S}(t)S^{\dagger}(\omega) \int_0^t e^{-i(\nu-\omega)\tau}d\tau\int_{-\infty}^{\infty} G_j(\nu) d\nu.
\end{equation}
Finally, using Eqs. \eqref{eqT1} and \eqref{second_term} one arrives at the master equation, 
\ba
\dot{\rho}_{\rm S}(t) &=& \mathcal{L}_j\left[\rho_{\rm S}(t) \right] \non\\
&=& \sum_{\omega}\tilde{\mathcal{R}}_j(\omega,t)  \mathcal{L}_{j, \omega}\left[\rho_{\rm S}(t) \right] + \text{h.c.};\non\\
\tilde{\mathcal{R}}_j(\omega,t) &\equiv& \int_{-\infty}^{\infty} d\nu G_j(\nu)\Big[\frac{\sin\left[\left(\nu - \omega \right)t\right]}{\nu - \omega} \non\\
&\pm& i \left(\frac{\cos\left[\left(\nu - \omega \right)t\right] - 1}{\nu - \omega}\right)\Big],
\label{rwaTsinc}
\ea
$\mathcal{L}_{j,\omega}[\rho_{\rm S}(t)]\equiv {\lambda_j(t)}^2[S^{\dagger}(\omega)S(\omega)\rho_{\rm S}(t)+S^{\dagger}(\omega) \rho_S(t)S(\omega)]$, and the h.c. denotes the hermitian conjugate.

\section{Master equation for a two-level system working medium}
\label{appB}
Now we focus on the  thermalization strokes by considering the dynamics of a two-level system coupled with a bath, via an interaction Hamiltonian $H_{\rm SB}$ with $S=\sigma_x$. During the first thermalization stroke $H_{\rm S}(t)=\frac{\omega_h}{2} \sigma_z$, while $H_{\rm S}(t) = \frac{\omega_{\rm c}}{2}\sigma_z$, during the second thermalization stroke. Hence, in general, in the interaction picture we can write,
\begin{equation}
\sigma_x(t)=e^{i\omega_{j} t}\sigma^++e^{-i\omega_{j} t}\sigma^-,
\end{equation}
where $j= \{\rm h,~c\}$, $\sigma^+=\frac{1}{2}(\sigma_x+i\sigma_y)$, and $\sigma^-=\frac{1}{2}(\sigma_x-i\sigma_y)$.
\begin{figure}[h]
\begin{center}
\includegraphics[height = 0.7\columnwidth, width = \columnwidth]{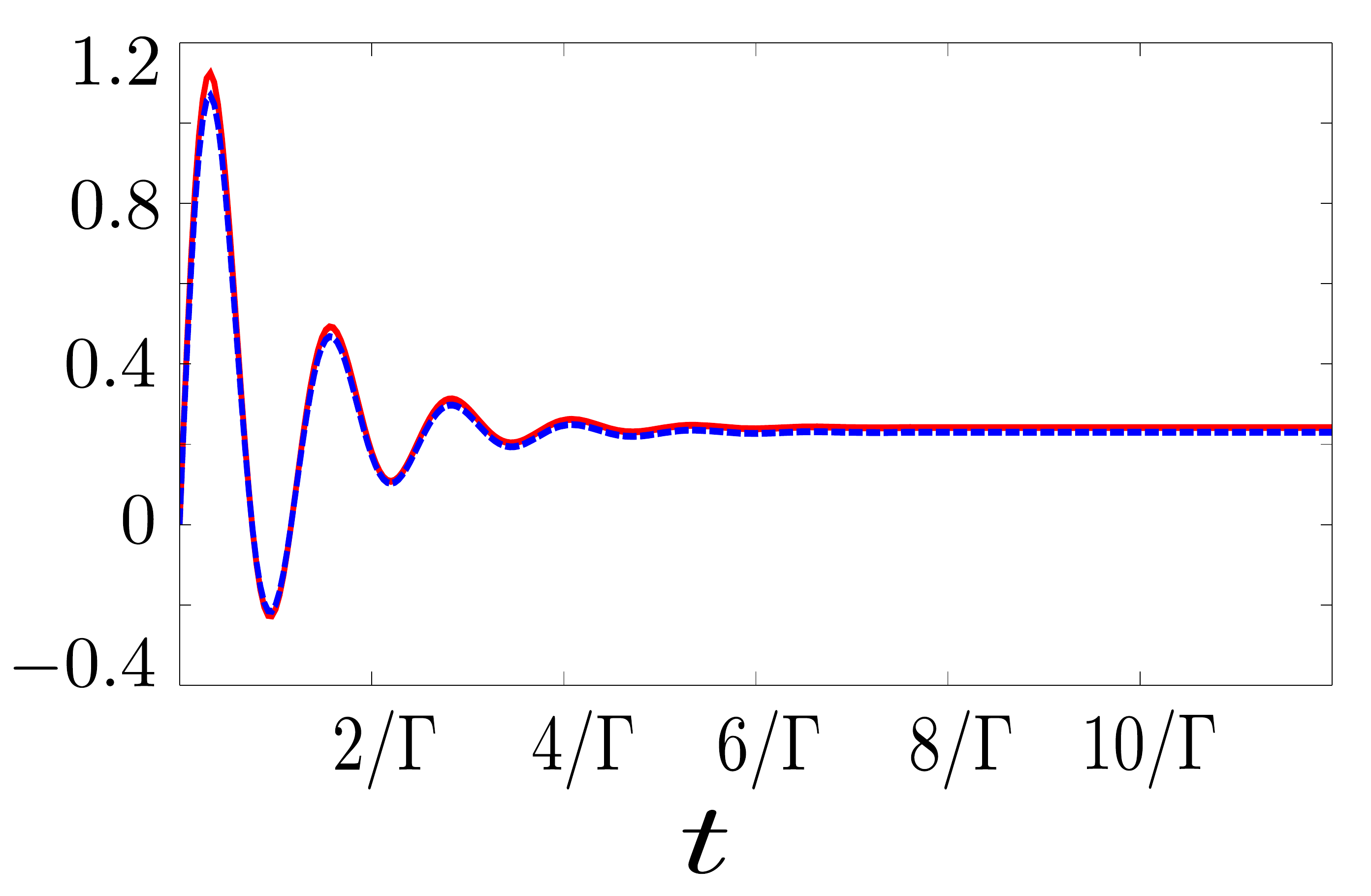}
\caption{Rapid intermittent coupling and decoupling between the WM and the thermal baths lead to $\mathcal{R}_{\rm h}(\omega_{\rm h},t)$ (red solid curve) and $\mathcal{R}_{\rm c}(\omega_{\rm c},t)$ (blue dashed curve)  becoming negative for intermediate times, thus resulting in non-Markovian dynamics of the WM. In this plot Lorentzian bath spectral function, Eq. (\ref{bath1}) has been used with $\delta=2$, $\Gamma=0.4$ and $\gamma_0 = 1$. Here, $\omega_j=100$, $\beta_j=0.0005$ for the hot ($\rm h$) bath and $\omega_j=80$, $\beta_j=0.01$ for the cold ($\rm c$) bath. }
\label{RjNM}
\end{center}
\end{figure}
Proceeding as before, we get the first term of the master equation \eqref{master_first1} as,
\begin{equation}
{\lambda_j(t)}^2[-\tilde{\mathcal{R}}_{j}(+\omega_{j},t)\sigma^-\sigma^+\rho_{\rm S}(t)-\tilde{\mathcal{R}}_{j}(-\omega_{j},t)\sigma^+\sigma^-\rho_{\rm S}(t)],
\end{equation}
where (see Fig. \ref{RjNM}),
\begin{eqnarray}
\label{expa}
\tilde{\mathcal{R}}_{j}(\pm\omega_{j},t) &=& \int_{-\infty}^{\infty}G(\nu)\int_0^t e^{i(\nu\mp\omega_{j})\tau} d\nu d\tau.
\end{eqnarray}
Similarly, evaluating the other terms, and considering a diagonal initial state, one arrives at the master equation (see Eqs. \eqref{rate}-\eqref{eqsinc}),
\begin{align}
\nonumber
&\dot{\rho}_{\rm S}(t)=\\
\nonumber
&2{\lambda_j(t)}^2 [\mathcal{R}_{j}(+\omega_{j},t)\sigma^+\rho_{\rm S}(t)\sigma^-+\mathcal{R}_{j}(-\omega_{j},t)\sigma^-\rho_{\rm S}(t)\sigma^+\\
&-\mathcal{R}_{j}(+\omega_{j},t)\sigma^-\sigma^+\rho_{\rm S}(t)-\mathcal{R}_{j}(-\omega_{j},t)\sigma^+\sigma^-\rho_{\rm S}(t)],
\label{master2}
\end{align} 
which finally leads us to the rate equations \eqref{rate}.

\section{Lorentzian bath spectral functions}
\label{appC}
We consider thermal baths with Lorentzian spectral functions, given by,
\begin{eqnarray}
\label{bath1}
\nonumber
&& G_{j}(\nu\geq 0)=\frac{\gamma_0\Gamma^2}{{(\nu-\omega_{j}-\delta)}^2+\Gamma^2},\\
&& G_{j}(\nu<0)=G_{j}(\nu\geq 0)e^{-\beta_{j}\nu},
\end{eqnarray}
\begin{figure}[H]
\begin{center}
\subfloat[]{\includegraphics[height = 0.6\columnwidth, width = \columnwidth]{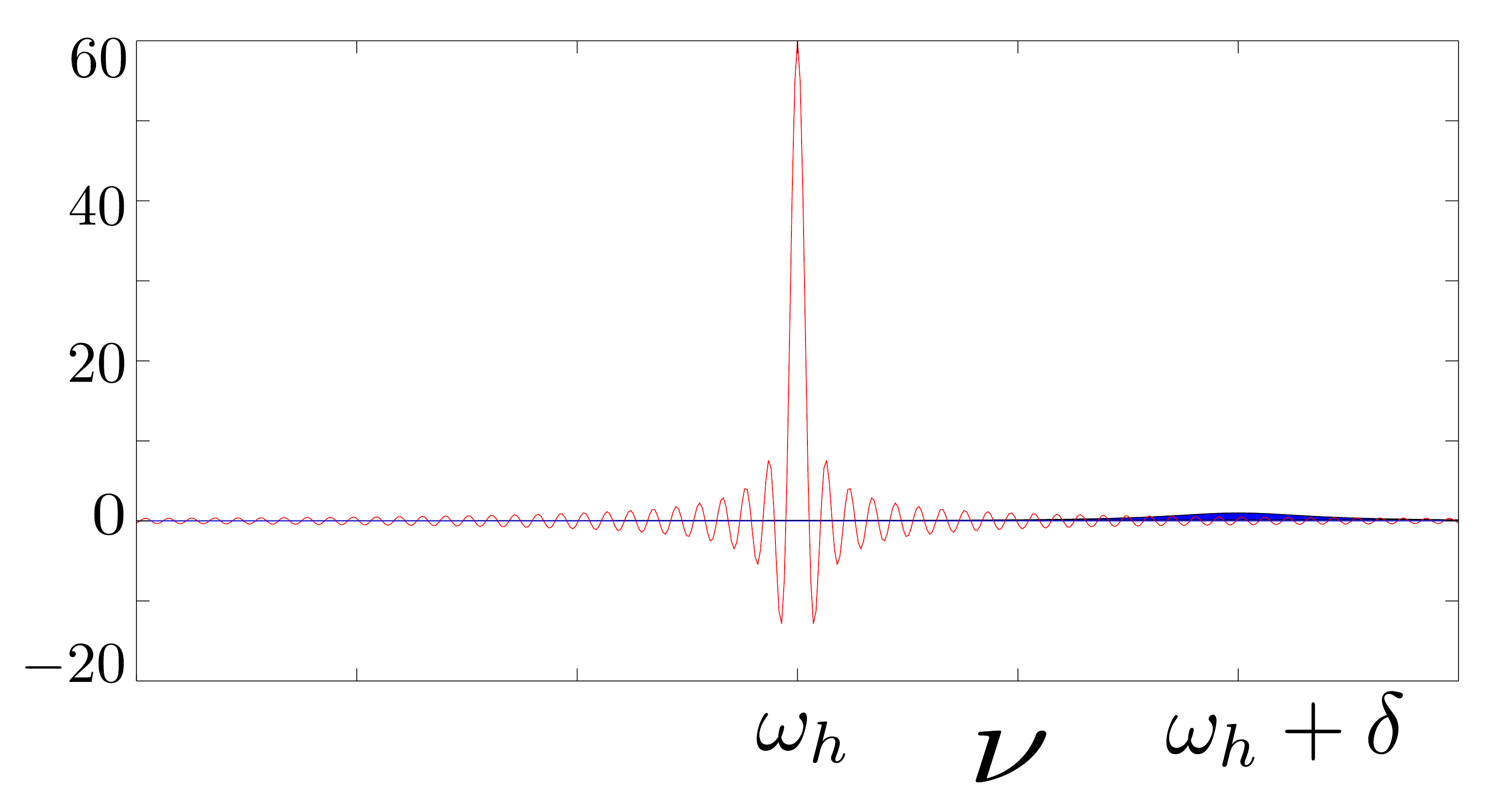}\label{new_markov_l}}
\hfill
\subfloat[]{\includegraphics[height = 0.6\columnwidth, width = \columnwidth]{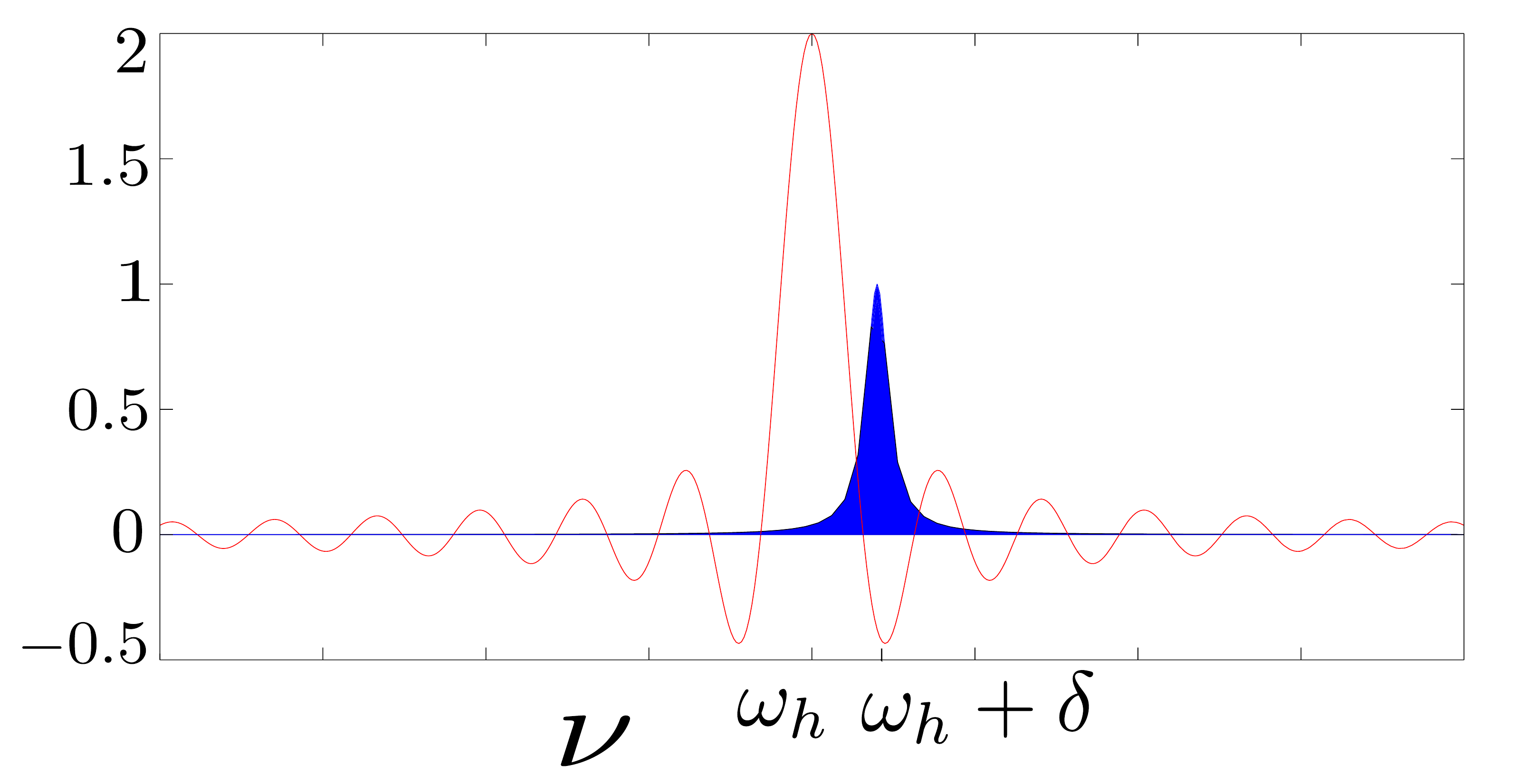}\label{new_markov_la}}
\caption{Blue filled curve shows the Lorentzian  hot bath spectral function given by Eq. (\ref{bath1}) and red one shows the function
$\sin([\nu-\omega_{\rm h}]t)/(\nu-\omega_{\rm h})=\sinc((\nu-\omega_{\rm h})t)t$, for $\Gamma=0.2$, $\gamma_0=1$, with $\delta=2$, $\omega_{\rm h}=100$, $\beta_{\rm h}=0.0005$. (a)  Markovian limit: $t=24/\Gamma$. (b) Anti-zeno limit: $t=0.8/\Gamma$.}
\label{Fig6whole}
\end{center}
\end{figure}
where $\gamma_0$ is the system-bath coupling strength, $\Gamma\sim 1/\tau_{\rm B}$ is the width of the spectrum and the bath spectral function shows a maximum at frequency $\omega_{\rm h,c}+\delta$.
As shown in Fig. \ref{new_markov_l}, the sinc function assumes the form of a delta function in the Markov limit $\tau_{\rm cp} \gg \tau_{\rm B}$, thus resulting in a 
vanishing overlap with the bath spectral function. On the other hand, larger overlap between the sinc function and the bath spectral functions in the anti-Zeno dynamics limit lead to enhanced heat flows (see Fig. \ref{new_markov_la}) and and faster thermalization (see Fig. \ref{therm_lorentz}).

\begin{figure}[H]
\begin{center}
\includegraphics[height = 0.7\columnwidth, width = \columnwidth]{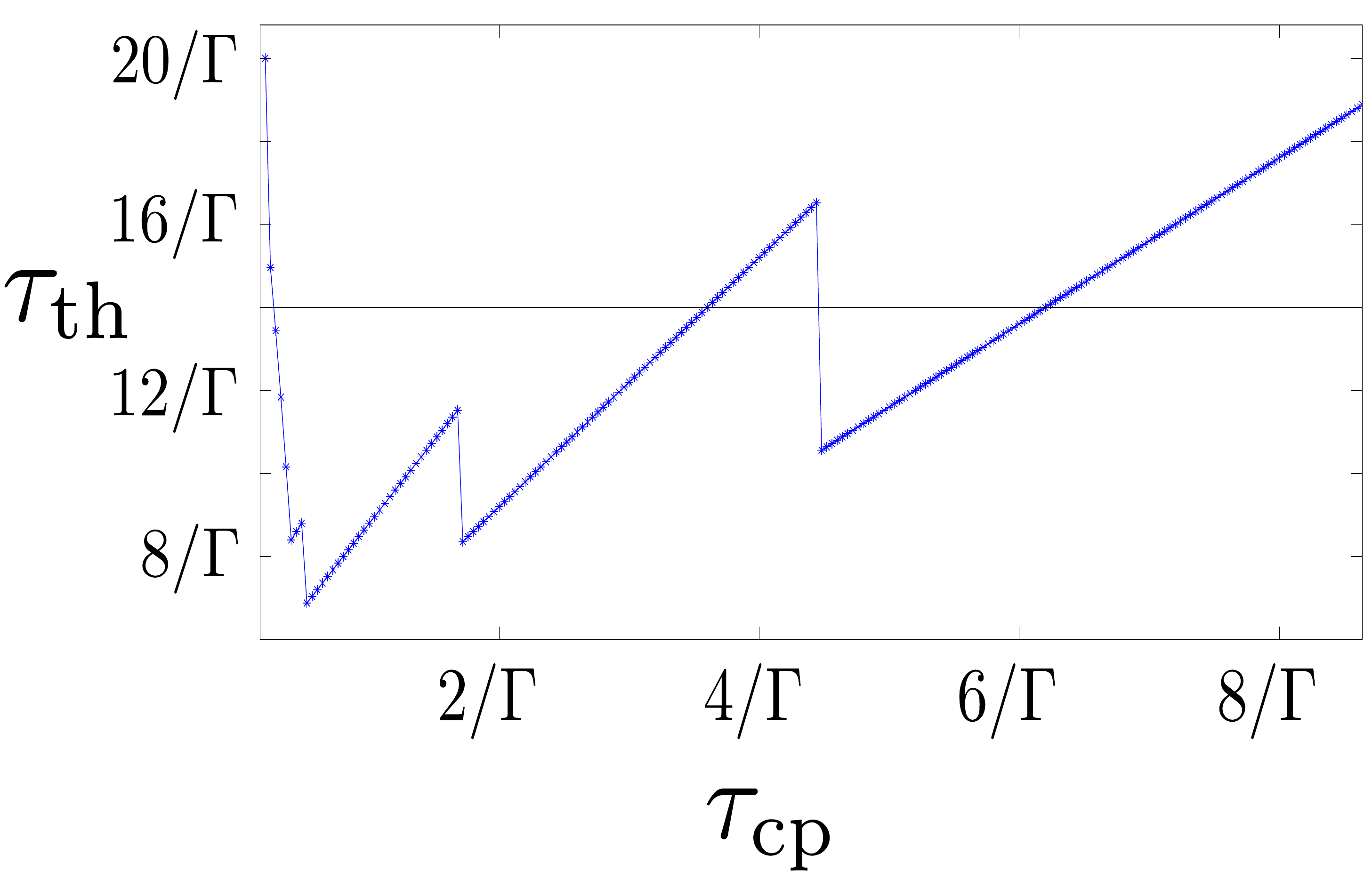}
\caption{Thermalization time for the first thermalization stroke of heat engine with Lorentzian bath spectral function, Eq. (\ref{bath1}) having $\delta=2$, $\Gamma=0.4$ and $\gamma_0 = 1$. Here  $\beta_{\rm h} =0.0005$, $\beta_{\rm c}=0.01$, $\omega_{\rm c}=80$, $\omega_{\rm h}=100$, $\epsilon < 0.0015$. $\tau_{\rm th}$ in the Markovian limit is shown by the black horizontal line.}
\label{therm_lorentz}
\end{center}
\end{figure}

\section{Super-ohmic bath spectral functions}
\label{appD}
We consider  Super-ohmic bath spectral functions, given by,
\begin{eqnarray}
\nonumber
G_{j}(\nu\geq 0) &=& \gamma_0\frac{\Theta(\nu-\omega_{j}+\delta){(\nu-\omega_{j}+\delta)}^s}{{(\bar{\nu})}^{s-1}}e^{\frac{-(\nu-\omega_{j}+\delta)}{\bar{\nu}}},\\
G_{j}(\nu< 0) &=& G_{j}(\nu\geq 0)e^{-\beta_{j}\nu}.
\label{bath2}
\end{eqnarray} 
\begin{figure}[H]
\begin{center}
\subfloat[]{\includegraphics[height = 0.6\columnwidth, width = \columnwidth]{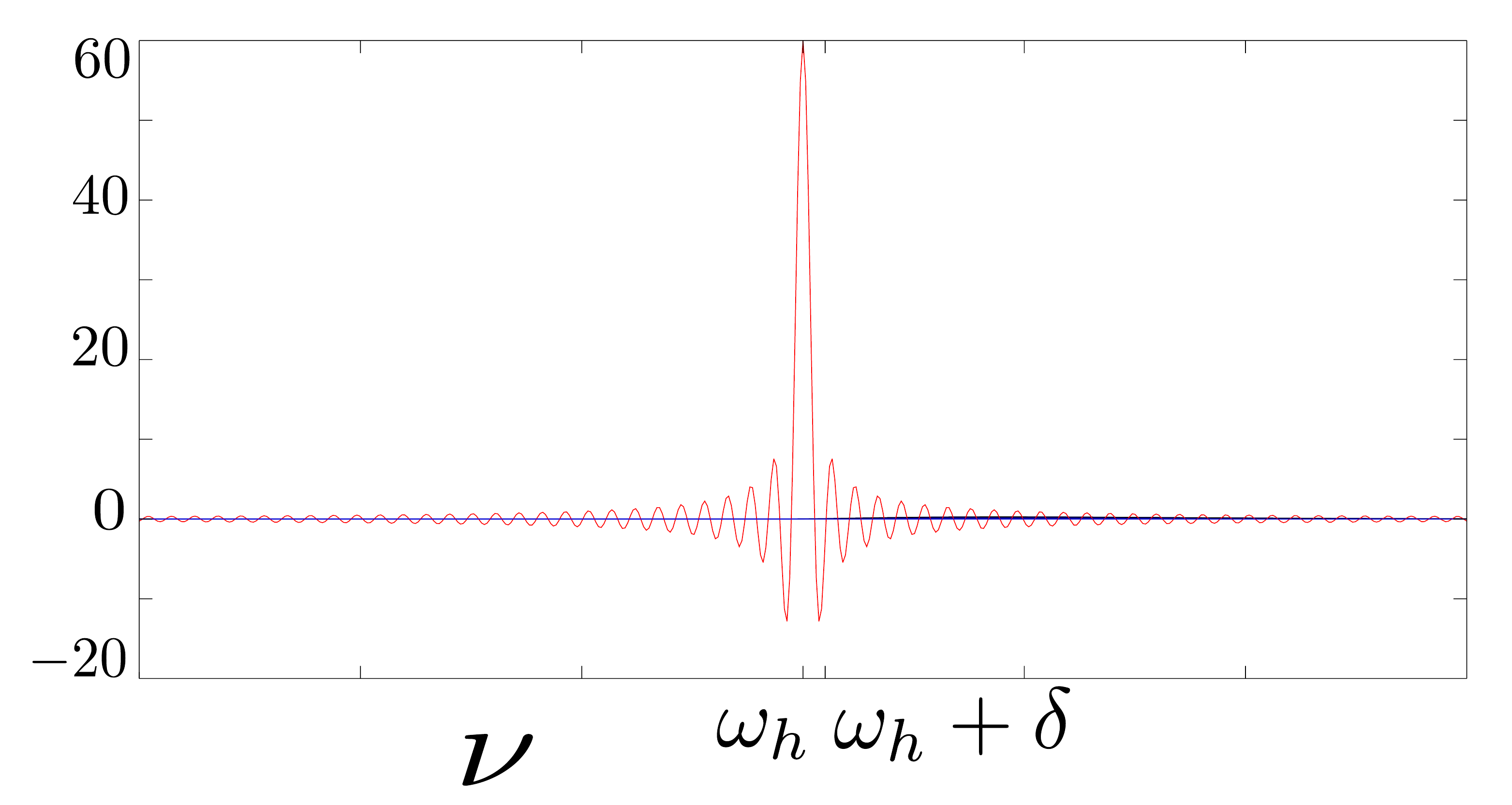}\label{new_mkv}}
\hfill
\subfloat[]{\includegraphics[height = 0.6\columnwidth, width = \columnwidth]{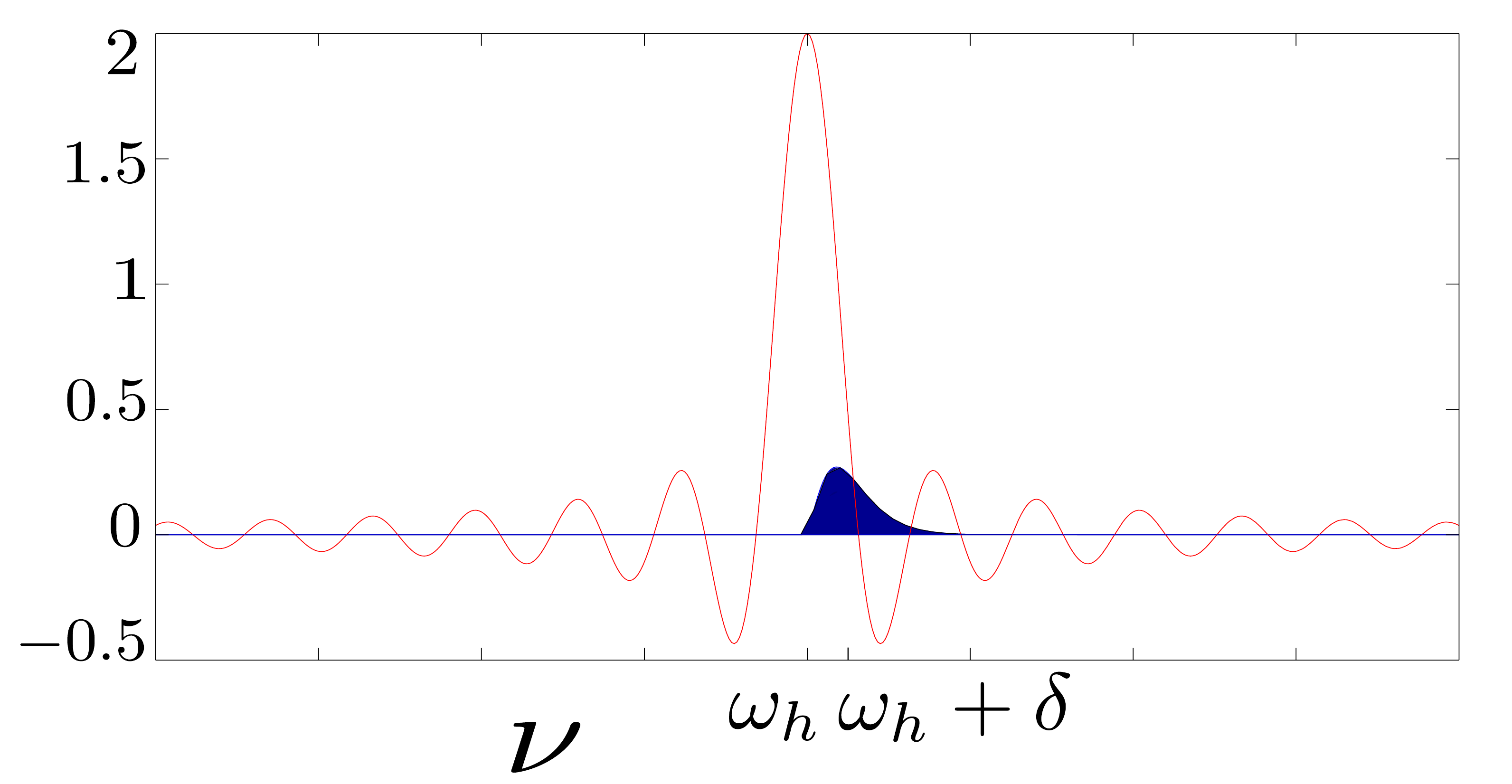}\label{new_azd}}
\caption{Blue filled curve shows the super-Ohmic hot bath spectral function  given by Eq. (\ref{bath1}) and red one shows the function $\sin([\nu-\omega_{\rm h}]t)/(\nu-\omega_{\rm h})=\sinc((\nu-\omega_{\rm h})t)t$, for  $\bar{\nu}=0.5$,  $\gamma_0=1$, with  $\delta=0.1$, $\omega_{\rm h}=100$, $\beta_{\rm h} = 0.0005$. (a) Markovian limit: $t=30/\bar{\nu}$. (b) Anti-zeno limit: $t=1/\bar{\nu}$.}
\label{Fig8whole}
\end{center}
\end{figure} 
Here $\bar{\nu} \sim 1/\tau_{\rm B}$, and as before, $\gamma_0$ is  the system-bath coupling strength. A small non-zero $\delta$ ensures that the bath spectral function and the sinc function attain maxima at different frequencies. We plot the bath spectral function and the sinc function for both the Markov and the anti-Zeno dynamics limits. 

\begin{figure}[H]
\begin{center}
\includegraphics[height = 0.7\columnwidth, width = \columnwidth]{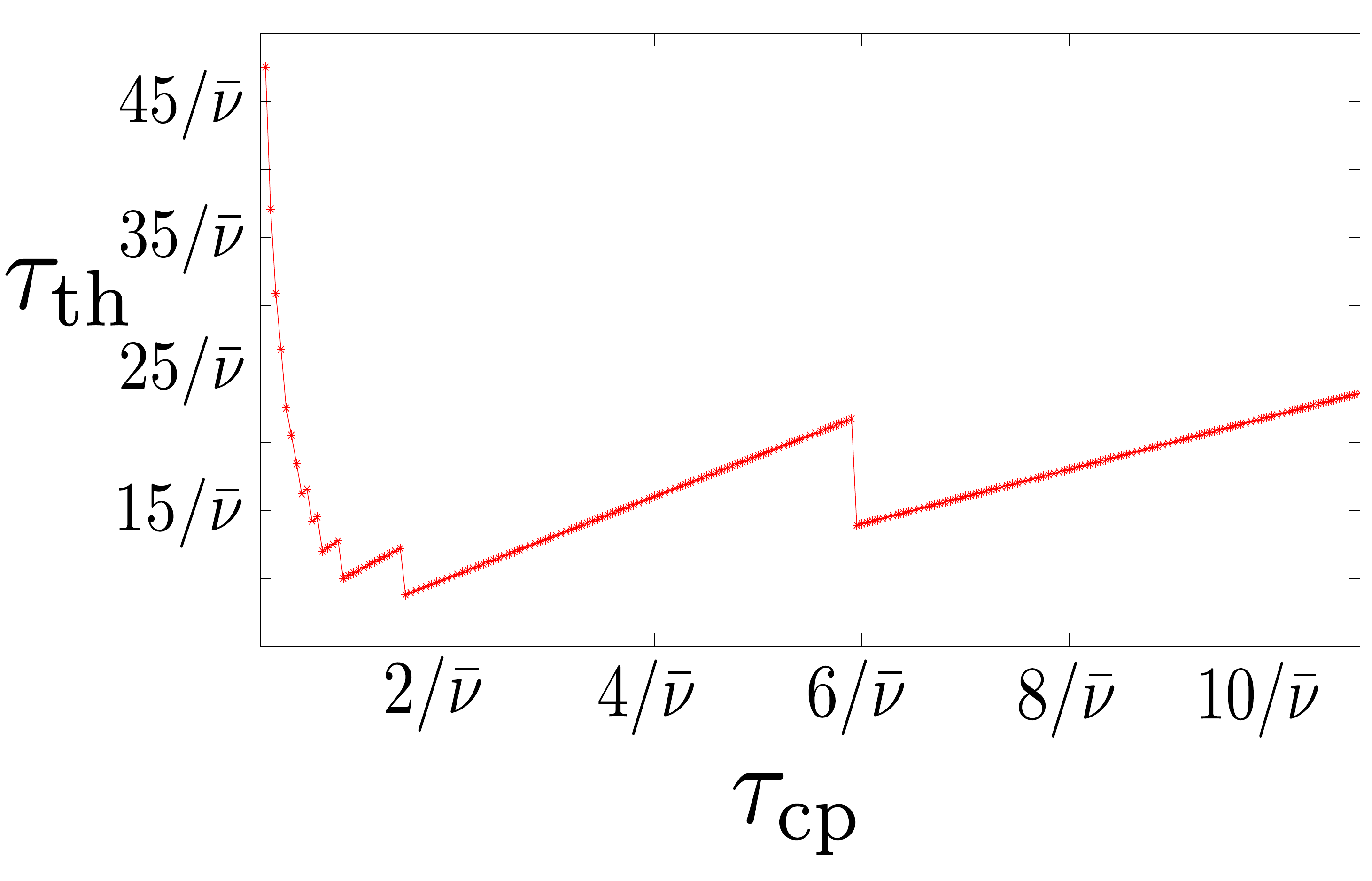}
\caption{Thermalization time for the first thermalization stroke of heat engine with super-Ohmic bath spectral function, Eq. (\ref{bath2}) having $\delta=0.1$, $\bar{\nu}=0.5$ and $\gamma_0 = 1$. Here  $\beta_{\rm h} =0.0005$, $\beta_{\rm c}=0.01$, $\omega_{\rm c}=80$, $\omega_{\rm h}=100$, $\epsilon < 0.0015$. $\tau_{\rm th}$ in the Markovian limit is shown by the black horizontal line.}
\label{therm_ohmic}
\end{center}
\end{figure}

As for the Lorentzian bath spectral functions, Figs. \ref{new_mkv} and \ref{new_azd} show significant overlap between the bath spectral function and the sinc function, only in the limit of anti-Zeno dynamics and a consequent faster thermalization (see Fig. \ref{therm_ohmic}).

\section*{Acknowledgements}
VM acknowledges Arnab Ghosh and Gershon Kurizki for careful reading of the manuscript, Science and Engineering Research Board (SERB) for Start-up Research Grant SRG/2019/000411 and IISER Berhampur for Seed grant.


\begin{thebibliography}{74}%
\makeatletter
\providecommand \@ifxundefined [1]{%
 \@ifx{#1\undefined}
}%
\providecommand \@ifnum [1]{%
 \ifnum #1\expandafter \@firstoftwo
 \else \expandafter \@secondoftwo
 \fi
}%
\providecommand \@ifx [1]{%
 \ifx #1\expandafter \@firstoftwo
 \else \expandafter \@secondoftwo
 \fi
}%
\providecommand \natexlab [1]{#1}%
\providecommand \enquote  [1]{``#1''}%
\providecommand \bibnamefont  [1]{#1}%
\providecommand \bibfnamefont [1]{#1}%
\providecommand \citenamefont [1]{#1}%
\providecommand \href@noop [0]{\@secondoftwo}%
\providecommand \href [0]{\begingroup \@sanitize@url \@href}%
\providecommand \@href[1]{\@@startlink{#1}\@@href}%
\providecommand \@@href[1]{\endgroup#1\@@endlink}%
\providecommand \@sanitize@url [0]{\catcode `\\12\catcode `\$12\catcode
  `\&12\catcode `\#12\catcode `\^12\catcode `\_12\catcode `\%12\relax}%
\providecommand \@@startlink[1]{}%
\providecommand \@@endlink[0]{}%
\providecommand \url  [0]{\begingroup\@sanitize@url \@url }%
\providecommand \@url [1]{\endgroup\@href {#1}{\urlprefix }}%
\providecommand \urlprefix  [0]{URL }%
\providecommand \Eprint [0]{\href }%
\providecommand \doibase [0]{http://dx.doi.org/}%
\providecommand \selectlanguage [0]{\@gobble}%
\providecommand \bibinfo  [0]{\@secondoftwo}%
\providecommand \bibfield  [0]{\@secondoftwo}%
\providecommand \translation [1]{[#1]}%
\providecommand \BibitemOpen [0]{}%
\providecommand \bibitemStop [0]{}%
\providecommand \bibitemNoStop [0]{.\EOS\space}%
\providecommand \EOS [0]{\spacefactor3000\relax}%
\providecommand \BibitemShut  [1]{\csname bibitem#1\endcsname}%
\let\auto@bib@innerbib\@empty
\bibitem [{\citenamefont {Golter}\ \emph {et~al.}(2016)\citenamefont {Golter},
  \citenamefont {Oo}, \citenamefont {Amezcua}, \citenamefont {Stewart},\ and\
  \citenamefont {Wang}}]{golter16optomechanical}%
  \BibitemOpen
  \bibfield  {author} {\bibinfo {author} {\bibfnamefont {D.~A.}\ \bibnamefont
  {Golter}}, \bibinfo {author} {\bibfnamefont {T.}~\bibnamefont {Oo}}, \bibinfo
  {author} {\bibfnamefont {M.}~\bibnamefont {Amezcua}}, \bibinfo {author}
  {\bibfnamefont {K.~A.}\ \bibnamefont {Stewart}}, \ and\ \bibinfo {author}
  {\bibfnamefont {H.}~\bibnamefont {Wang}},\ }\href {\doibase
  10.1103/PhysRevLett.116.143602} {\bibfield  {journal} {\bibinfo  {journal}
  {Phys. Rev. Lett.}\ }\textbf {\bibinfo {volume} {116}},\ \bibinfo {pages}
  {143602} (\bibinfo {year} {2016})}\BibitemShut {NoStop}%
\bibitem [{\citenamefont {Accanto}\ \emph {et~al.}(2017)\citenamefont
  {Accanto}, \citenamefont {de~Roque}, \citenamefont {Galvan-Sosa},
  \citenamefont {Christodoulou}, \citenamefont {Moreels},\ and\ \citenamefont
  {van Hulst}}]{accanto17rapid}%
  \BibitemOpen
  \bibfield  {author} {\bibinfo {author} {\bibfnamefont {N.}~\bibnamefont
  {Accanto}}, \bibinfo {author} {\bibfnamefont {P.~M.}\ \bibnamefont
  {de~Roque}}, \bibinfo {author} {\bibfnamefont {M.}~\bibnamefont
  {Galvan-Sosa}}, \bibinfo {author} {\bibfnamefont {S.}~\bibnamefont
  {Christodoulou}}, \bibinfo {author} {\bibfnamefont {I.}~\bibnamefont
  {Moreels}}, \ and\ \bibinfo {author} {\bibfnamefont {N.~F.}\ \bibnamefont
  {van Hulst}},\ }\href {\doibase 10.1038/lsa.2016.239} {\bibfield  {journal}
  {\bibinfo  {journal} {Light: Science \& Applications}\ }\textbf {\bibinfo
  {volume} {6}},\ \bibinfo {pages} {e16239} (\bibinfo {year}
  {2017})}\BibitemShut {NoStop}%
\bibitem [{\citenamefont {Perreault}\ \emph {et~al.}(2017)\citenamefont
  {Perreault}, \citenamefont {Mukherjee},\ and\ \citenamefont
  {Zare}}]{perreault17quantum}%
  \BibitemOpen
  \bibfield  {author} {\bibinfo {author} {\bibfnamefont {W.~E.}\ \bibnamefont
  {Perreault}}, \bibinfo {author} {\bibfnamefont {N.}~\bibnamefont
  {Mukherjee}}, \ and\ \bibinfo {author} {\bibfnamefont {R.~N.}\ \bibnamefont
  {Zare}},\ }\href {\doibase 10.1126/science.aao3116} {\bibfield  {journal}
  {\bibinfo  {journal} {Science}\ }\textbf {\bibinfo {volume} {358}},\ \bibinfo
  {pages} {356} (\bibinfo {year} {2017})}\BibitemShut {NoStop}%
\bibitem [{\citenamefont {Rossi}\ \emph {et~al.}(2018)\citenamefont {Rossi},
  \citenamefont {Mason}, \citenamefont {Chen}, \citenamefont {Tsaturyan},\ and\
  \citenamefont {Schliesser}}]{rossi18measurement}%
  \BibitemOpen
  \bibfield  {author} {\bibinfo {author} {\bibfnamefont {M.}~\bibnamefont
  {Rossi}}, \bibinfo {author} {\bibfnamefont {D.}~\bibnamefont {Mason}},
  \bibinfo {author} {\bibfnamefont {J.}~\bibnamefont {Chen}}, \bibinfo {author}
  {\bibfnamefont {Y.}~\bibnamefont {Tsaturyan}}, \ and\ \bibinfo {author}
  {\bibfnamefont {A.}~\bibnamefont {Schliesser}},\ }\href {\doibase
  10.1038/s41586-018-0643-8} {\bibfield  {journal} {\bibinfo  {journal}
  {Nature}\ }\textbf {\bibinfo {volume} {563}},\ \bibinfo {pages} {53}
  (\bibinfo {year} {2018})}\BibitemShut {NoStop}%
\bibitem [{\citenamefont {Giovannetti}\ \emph {et~al.}(2011)\citenamefont
  {Giovannetti}, \citenamefont {Lloyd},\ and\ \citenamefont
  {Maccone}}]{giovannetti11advances}%
  \BibitemOpen
  \bibfield  {author} {\bibinfo {author} {\bibfnamefont {V.}~\bibnamefont
  {Giovannetti}}, \bibinfo {author} {\bibfnamefont {S.}~\bibnamefont {Lloyd}},
  \ and\ \bibinfo {author} {\bibfnamefont {L.}~\bibnamefont {Maccone}},\ }\href
  {\doibase 10.1038/nphoton.2011.35} {\bibfield  {journal} {\bibinfo  {journal}
  {Nature Photonics}\ }\textbf {\bibinfo {volume} {5}},\ \bibinfo {pages} {222}
  (\bibinfo {year} {2011})}\BibitemShut {NoStop}%
\bibitem [{\citenamefont {Kurizki}\ \emph {et~al.}(2015)\citenamefont
  {Kurizki}, \citenamefont {Bertet}, \citenamefont {Kubo}, \citenamefont
  {M{\o}lmer}, \citenamefont {Petrosyan}, \citenamefont {Rabl},\ and\
  \citenamefont {Schmiedmayer}}]{kurizki15quantum}%
  \BibitemOpen
  \bibfield  {author} {\bibinfo {author} {\bibfnamefont {G.}~\bibnamefont
  {Kurizki}}, \bibinfo {author} {\bibfnamefont {P.}~\bibnamefont {Bertet}},
  \bibinfo {author} {\bibfnamefont {Y.}~\bibnamefont {Kubo}}, \bibinfo {author}
  {\bibfnamefont {K.}~\bibnamefont {M{\o}lmer}}, \bibinfo {author}
  {\bibfnamefont {D.}~\bibnamefont {Petrosyan}}, \bibinfo {author}
  {\bibfnamefont {P.}~\bibnamefont {Rabl}}, \ and\ \bibinfo {author}
  {\bibfnamefont {J.}~\bibnamefont {Schmiedmayer}},\ }\href {\doibase
  10.1073/pnas.1419326112} {\bibfield  {journal} {\bibinfo  {journal}
  {Proceedings of the National Academy of Sciences}\ }\textbf {\bibinfo
  {volume} {112}},\ \bibinfo {pages} {3866} (\bibinfo {year}
  {2015})}\BibitemShut {NoStop}%
\bibitem [{\citenamefont {Brantut}\ \emph {et~al.}(2013)\citenamefont
  {Brantut}, \citenamefont {Grenier}, \citenamefont {Meineke}, \citenamefont
  {Stadler}, \citenamefont {Krinner}, \citenamefont {Kollath}, \citenamefont
  {Esslinger},\ and\ \citenamefont {Georges}}]{brantut13a}%
  \BibitemOpen
  \bibfield  {author} {\bibinfo {author} {\bibfnamefont {J.-P.}\ \bibnamefont
  {Brantut}}, \bibinfo {author} {\bibfnamefont {C.}~\bibnamefont {Grenier}},
  \bibinfo {author} {\bibfnamefont {J.}~\bibnamefont {Meineke}}, \bibinfo
  {author} {\bibfnamefont {D.}~\bibnamefont {Stadler}}, \bibinfo {author}
  {\bibfnamefont {S.}~\bibnamefont {Krinner}}, \bibinfo {author} {\bibfnamefont
  {C.}~\bibnamefont {Kollath}}, \bibinfo {author} {\bibfnamefont
  {T.}~\bibnamefont {Esslinger}}, \ and\ \bibinfo {author} {\bibfnamefont
  {A.}~\bibnamefont {Georges}},\ }\href {\doibase 10.1126/science.1242308}
  {\bibfield  {journal} {\bibinfo  {journal} {Science}\ }\textbf {\bibinfo
  {volume} {342}},\ \bibinfo {pages} {713} (\bibinfo {year}
  {2013})}\BibitemShut {NoStop}%
\bibitem [{\citenamefont {Bernien}\ \emph {et~al.}(2017)\citenamefont
  {Bernien}, \citenamefont {Schwartz}, \citenamefont {Keesling}, \citenamefont
  {Levine}, \citenamefont {Omran}, \citenamefont {Pichler}, \citenamefont
  {Choi}, \citenamefont {Zibrov}, \citenamefont {Endres}, \citenamefont
  {Greiner}, \citenamefont {Vuletic},\ and\ \citenamefont
  {Lukin}}]{bernien17probing}%
  \BibitemOpen
  \bibfield  {author} {\bibinfo {author} {\bibfnamefont {H.}~\bibnamefont
  {Bernien}}, \bibinfo {author} {\bibfnamefont {S.}~\bibnamefont {Schwartz}},
  \bibinfo {author} {\bibfnamefont {A.}~\bibnamefont {Keesling}}, \bibinfo
  {author} {\bibfnamefont {H.}~\bibnamefont {Levine}}, \bibinfo {author}
  {\bibfnamefont {A.}~\bibnamefont {Omran}}, \bibinfo {author} {\bibfnamefont
  {H.}~\bibnamefont {Pichler}}, \bibinfo {author} {\bibfnamefont
  {S.}~\bibnamefont {Choi}}, \bibinfo {author} {\bibfnamefont {A.~S.}\
  \bibnamefont {Zibrov}}, \bibinfo {author} {\bibfnamefont {M.}~\bibnamefont
  {Endres}}, \bibinfo {author} {\bibfnamefont {M.}~\bibnamefont {Greiner}},
  \bibinfo {author} {\bibfnamefont {V.}~\bibnamefont {Vuletic}}, \ and\
  \bibinfo {author} {\bibfnamefont {M.~D.}\ \bibnamefont {Lukin}},\ }\href
  {\doibase 10.1038/nature24622} {\bibfield  {journal} {\bibinfo  {journal}
  {Nature}\ }\textbf {\bibinfo {volume} {551}},\ \bibinfo {pages} {579}
  (\bibinfo {year} {2017})}\BibitemShut {NoStop}%
\bibitem [{\citenamefont {Zhang}\ \emph {et~al.}(2017)\citenamefont {Zhang},
  \citenamefont {Pagano}, \citenamefont {Hess}, \citenamefont {Kyprianidis},
  \citenamefont {Becker}, \citenamefont {Kaplan}, \citenamefont {Gorshkov},
  \citenamefont {Gong},\ and\ \citenamefont {Monroe}}]{zhang17observation}%
  \BibitemOpen
  \bibfield  {author} {\bibinfo {author} {\bibfnamefont {J.}~\bibnamefont
  {Zhang}}, \bibinfo {author} {\bibfnamefont {G.}~\bibnamefont {Pagano}},
  \bibinfo {author} {\bibfnamefont {P.~W.}\ \bibnamefont {Hess}}, \bibinfo
  {author} {\bibfnamefont {A.}~\bibnamefont {Kyprianidis}}, \bibinfo {author}
  {\bibfnamefont {P.}~\bibnamefont {Becker}}, \bibinfo {author} {\bibfnamefont
  {H.}~\bibnamefont {Kaplan}}, \bibinfo {author} {\bibfnamefont {A.~V.}\
  \bibnamefont {Gorshkov}}, \bibinfo {author} {\bibfnamefont {Z.-X.}\
  \bibnamefont {Gong}}, \ and\ \bibinfo {author} {\bibfnamefont
  {C.}~\bibnamefont {Monroe}},\ }\href {\doibase 10.1038/nature24654}
  {\bibfield  {journal} {\bibinfo  {journal} {Nature}\ }\textbf {\bibinfo
  {volume} {551}},\ \bibinfo {pages} {601} (\bibinfo {year}
  {2017})}\BibitemShut {NoStop}%
\bibitem [{\citenamefont {Klatzow}\ \emph {et~al.}(2019)\citenamefont
  {Klatzow}, \citenamefont {Becker}, \citenamefont {Ledingham}, \citenamefont
  {Weinzetl}, \citenamefont {Kaczmarek}, \citenamefont {Saunders},
  \citenamefont {Nunn}, \citenamefont {Walmsley}, \citenamefont {Uzdin},\ and\
  \citenamefont {Poem}}]{klatzow19experimental}%
  \BibitemOpen
  \bibfield  {author} {\bibinfo {author} {\bibfnamefont {J.}~\bibnamefont
  {Klatzow}}, \bibinfo {author} {\bibfnamefont {J.~N.}\ \bibnamefont {Becker}},
  \bibinfo {author} {\bibfnamefont {P.~M.}\ \bibnamefont {Ledingham}}, \bibinfo
  {author} {\bibfnamefont {C.}~\bibnamefont {Weinzetl}}, \bibinfo {author}
  {\bibfnamefont {K.~T.}\ \bibnamefont {Kaczmarek}}, \bibinfo {author}
  {\bibfnamefont {D.~J.}\ \bibnamefont {Saunders}}, \bibinfo {author}
  {\bibfnamefont {J.}~\bibnamefont {Nunn}}, \bibinfo {author} {\bibfnamefont
  {I.~A.}\ \bibnamefont {Walmsley}}, \bibinfo {author} {\bibfnamefont
  {R.}~\bibnamefont {Uzdin}}, \ and\ \bibinfo {author} {\bibfnamefont
  {E.}~\bibnamefont {Poem}},\ }\href {\doibase 10.1103/PhysRevLett.122.110601}
  {\bibfield  {journal} {\bibinfo  {journal} {Phys. Rev. Lett.}\ }\textbf
  {\bibinfo {volume} {122}},\ \bibinfo {pages} {110601} (\bibinfo {year}
  {2019})}\BibitemShut {NoStop}%
\bibitem [{\citenamefont {Peterson}\ \emph {et~al.}(2019)\citenamefont
  {Peterson}, \citenamefont {Batalh\~ao}, \citenamefont {Herrera},
  \citenamefont {Souza}, \citenamefont {Sarthour}, \citenamefont {Oliveira},\
  and\ \citenamefont {Serra}}]{peterson19experimental}%
  \BibitemOpen
  \bibfield  {author} {\bibinfo {author} {\bibfnamefont {J.~P.~S.}\
  \bibnamefont {Peterson}}, \bibinfo {author} {\bibfnamefont {T.~B.}\
  \bibnamefont {Batalh\~ao}}, \bibinfo {author} {\bibfnamefont
  {M.}~\bibnamefont {Herrera}}, \bibinfo {author} {\bibfnamefont {A.~M.}\
  \bibnamefont {Souza}}, \bibinfo {author} {\bibfnamefont {R.~S.}\ \bibnamefont
  {Sarthour}}, \bibinfo {author} {\bibfnamefont {I.~S.}\ \bibnamefont
  {Oliveira}}, \ and\ \bibinfo {author} {\bibfnamefont {R.~M.}\ \bibnamefont
  {Serra}},\ }\href {\doibase 10.1103/PhysRevLett.123.240601} {\bibfield
  {journal} {\bibinfo  {journal} {Phys. Rev. Lett.}\ }\textbf {\bibinfo
  {volume} {123}},\ \bibinfo {pages} {240601} (\bibinfo {year}
  {2019})}\BibitemShut {NoStop}%
\bibitem [{\citenamefont {Kosloff}(2013)}]{kosloff13quantum}%
  \BibitemOpen
  \bibfield  {author} {\bibinfo {author} {\bibfnamefont {R.}~\bibnamefont
  {Kosloff}},\ }\href {\doibase 10.3390/e15062100} {\bibfield  {journal}
  {\bibinfo  {journal} {Entropy}\ }\textbf {\bibinfo {volume} {15}},\ \bibinfo
  {pages} {2100} (\bibinfo {year} {2013})}\BibitemShut {NoStop}%
\bibitem [{\citenamefont {Gelbwaser-Klimovsky}\ \emph
  {et~al.}(2015)\citenamefont {Gelbwaser-Klimovsky}, \citenamefont {Niedenzu},\
  and\ \citenamefont {Kurizki}}]{klimovsky15thermodynamics}%
  \BibitemOpen
  \bibfield  {author} {\bibinfo {author} {\bibfnamefont {D.}~\bibnamefont
  {Gelbwaser-Klimovsky}}, \bibinfo {author} {\bibfnamefont {W.}~\bibnamefont
  {Niedenzu}}, \ and\ \bibinfo {author} {\bibfnamefont {G.}~\bibnamefont
  {Kurizki}},\ }\href {\doibase http://dx.doi.org/10.1016/bs.aamop.2015.07.002}
  {\bibfield  {journal} {\bibinfo  {journal} {Advances In Atomic, Molecular,
  and Optical Physics}\ }\textbf {\bibinfo {volume} {64}},\ \bibinfo {pages}
  {329 } (\bibinfo {year} {2015})}\BibitemShut {NoStop}%
\bibitem [{\citenamefont {Vinjanampathy}\ and\ \citenamefont
  {Anders}(2016)}]{vinjanampathy16quantum}%
  \BibitemOpen
  \bibfield  {author} {\bibinfo {author} {\bibfnamefont {S.}~\bibnamefont
  {Vinjanampathy}}\ and\ \bibinfo {author} {\bibfnamefont {J.}~\bibnamefont
  {Anders}},\ }\href {\doibase 10.1080/00107514.2016.1201896} {\bibfield
  {journal} {\bibinfo  {journal} {Contemporary Physics}\ }\textbf {\bibinfo
  {volume} {57}},\ \bibinfo {pages} {545} (\bibinfo {year} {2016})}\BibitemShut
  {NoStop}%
\bibitem [{\citenamefont {Alicki}\ and\ \citenamefont
  {Kosloff}(2019)}]{alicki18introduction}%
  \BibitemOpen
  \bibfield  {author} {\bibinfo {author} {\bibfnamefont {R.}~\bibnamefont
  {Alicki}}\ and\ \bibinfo {author} {\bibfnamefont {R.}~\bibnamefont
  {Kosloff}},\ }\href {\doibase https://doi.org/10.1007/978-3-319-99046-0_1}
  {\emph {\bibinfo {title} {Introduction to Quantum Thermodynamics: History and
  Prospects. Thermodynamics in the Quantum Regime}}},\ edited by\ \bibinfo
  {editor} {\bibfnamefont {F.}~\bibnamefont {Binder}}, \bibinfo {editor}
  {\bibfnamefont {L.~A.}\ \bibnamefont {Correa}}, \bibinfo {editor}
  {\bibfnamefont {C.}~\bibnamefont {Gogolin}}, \bibinfo {editor} {\bibfnamefont
  {J.}~\bibnamefont {Anders}}, \ and\ \bibinfo {editor} {\bibfnamefont
  {G.}~\bibnamefont {Adesso}}\ (\bibinfo  {publisher} {Springer, Cham},\
  \bibinfo {year} {2019}), pp. 1-33\BibitemShut {NoStop}%
\bibitem [{\citenamefont {Binder}\ \emph {et~al.}(2018)\citenamefont {Binder},
  \citenamefont {Correa}, \citenamefont {Gogolin}, \citenamefont {Anders},\
  and\ \citenamefont {Adesso}}]{binder18book}%
  \BibitemOpen
  \bibinfo {editor} {\bibfnamefont {F.}~\bibnamefont {Binder}}, \bibinfo
  {editor} {\bibfnamefont {L.~A.}\ \bibnamefont {Correa}}, \bibinfo {editor}
  {\bibfnamefont {C.}~\bibnamefont {Gogolin}}, \bibinfo {editor} {\bibfnamefont
  {J.}~\bibnamefont {Anders}}, \ and\ \bibinfo {editor} {\bibfnamefont
  {G.}~\bibnamefont {Adesso}},\ eds.,\ \href@noop {} {\emph {\bibinfo {title}
  {Thermodynamics in the quantum regime}}}\ (\bibinfo  {publisher} {Springer
  International Publishing},\ \bibinfo {year} {2018})\BibitemShut {NoStop}%
\bibitem [{\citenamefont {Alicki}(1979)}]{alicki79the}%
  \BibitemOpen
  \bibfield  {author} {\bibinfo {author} {\bibfnamefont {R.}~\bibnamefont
  {Alicki}},\ }\href {\doibase https://doi.org/10.1088/0305-4470/12/5/007}
  {\bibfield  {journal} {\bibinfo  {journal} {Journal of Physics A:
  Mathematical and General}\ }\textbf {\bibinfo {volume} {12}},\ \bibinfo
  {pages} {L103} (\bibinfo {year} {1979})}\BibitemShut {NoStop}%
\bibitem [{\citenamefont {Gelbwaser-Klimovsky}\ \emph
  {et~al.}(2013{\natexlab{a}})\citenamefont {Gelbwaser-Klimovsky},
  \citenamefont {Alicki},\ and\ \citenamefont {Kurizki}}]{klimovsky13minimal}%
  \BibitemOpen
  \bibfield  {author} {\bibinfo {author} {\bibfnamefont {D.}~\bibnamefont
  {Gelbwaser-Klimovsky}}, \bibinfo {author} {\bibfnamefont {R.}~\bibnamefont
  {Alicki}}, \ and\ \bibinfo {author} {\bibfnamefont {G.}~\bibnamefont
  {Kurizki}},\ }\href {\doibase 10.1103/PhysRevE.87.012140} {\bibfield
  {journal} {\bibinfo  {journal} {Phys. Rev. E}\ }\textbf {\bibinfo {volume}
  {87}},\ \bibinfo {pages} {012140} (\bibinfo {year}
  {2013}{\natexlab{a}})}\BibitemShut {NoStop}%
\bibitem [{\citenamefont {Alicki}(2014)}]{alicki14quantum}%
  \BibitemOpen
  \bibfield  {author} {\bibinfo {author} {\bibfnamefont {R.}~\bibnamefont
  {Alicki}},\ }\href {\doibase https://doi.org/10.1142/S1230161214400022}
  {\bibfield  {journal} {\bibinfo  {journal} {Open Systems And Information
  Dynamics}\ }\textbf {\bibinfo {volume} {21}},\ \bibinfo {pages} {1440002}
  (\bibinfo {year} {2014})}\BibitemShut {NoStop}%
\bibitem [{\citenamefont {Ro\ss{}nagel}\ \emph {et~al.}(2014)\citenamefont
  {Ro\ss{}nagel}, \citenamefont {Abah}, \citenamefont {Schmidt-Kaler},
  \citenamefont {Singer},\ and\ \citenamefont {Lutz}}]{rossnage14nanoscale}%
  \BibitemOpen
  \bibfield  {author} {\bibinfo {author} {\bibfnamefont {J.}~\bibnamefont
  {Ro\ss{}nagel}}, \bibinfo {author} {\bibfnamefont {O.}~\bibnamefont {Abah}},
  \bibinfo {author} {\bibfnamefont {F.}~\bibnamefont {Schmidt-Kaler}}, \bibinfo
  {author} {\bibfnamefont {K.}~\bibnamefont {Singer}}, \ and\ \bibinfo {author}
  {\bibfnamefont {E.}~\bibnamefont {Lutz}},\ }\href {\doibase
  10.1103/PhysRevLett.112.030602} {\bibfield  {journal} {\bibinfo  {journal}
  {Phys. Rev. Lett.}\ }\textbf {\bibinfo {volume} {112}},\ \bibinfo {pages}
  {030602} (\bibinfo {year} {2014})}\BibitemShut {NoStop}%
\bibitem [{\citenamefont {Kosloff}\ and\ \citenamefont
  {Rezek}(2017)}]{kosloff17the}%
  \BibitemOpen
  \bibfield  {author} {\bibinfo {author} {\bibfnamefont {R.}~\bibnamefont
  {Kosloff}}\ and\ \bibinfo {author} {\bibfnamefont {Y.}~\bibnamefont
  {Rezek}},\ }\href {\doibase https://doi.org/10.3390/e19040136} {\bibfield
  {journal} {\bibinfo  {journal} {Entropy}\ }\textbf {\bibinfo {volume} {19}},\
  \bibinfo {pages} {136} (\bibinfo {year} {2017})}\BibitemShut {NoStop}%
\bibitem [{\citenamefont {Ghosh}\ \emph {et~al.}(2017)\citenamefont {Ghosh},
  \citenamefont {Latune}, \citenamefont {Davidovich},\ and\ \citenamefont
  {Kurizki}}]{ghosh17catalysis}%
  \BibitemOpen
  \bibfield  {author} {\bibinfo {author} {\bibfnamefont {A.}~\bibnamefont
  {Ghosh}}, \bibinfo {author} {\bibfnamefont {C.~L.}\ \bibnamefont {Latune}},
  \bibinfo {author} {\bibfnamefont {L.}~\bibnamefont {Davidovich}}, \ and\
  \bibinfo {author} {\bibfnamefont {G.}~\bibnamefont {Kurizki}},\ }\href
  {\doibase 10.1073/pnas.1711381114} {\bibfield  {journal} {\bibinfo  {journal}
  {Proceedings of the National Academy of Sciences}\ }\textbf {\bibinfo
  {volume} {114}},\ \bibinfo {pages} {12156} (\bibinfo {year}
  {2017})}\BibitemShut {NoStop}%
\bibitem [{\citenamefont {Campisi}\ and\ \citenamefont
  {Fazio}(2016)}]{campisi16the}%
  \BibitemOpen
  \bibfield  {author} {\bibinfo {author} {\bibfnamefont {M.}~\bibnamefont
  {Campisi}}\ and\ \bibinfo {author} {\bibfnamefont {R.}~\bibnamefont
  {Fazio}},\ }\href {\doibase https://doi.org/10.1038/ncomms11895} {\bibfield
  {journal} {\bibinfo  {journal} {Nat. Comm.}\ }\textbf {\bibinfo {volume}
  {7}},\ \bibinfo {pages} {11895} (\bibinfo {year} {2016})}\BibitemShut
  {NoStop}%
\bibitem [{\citenamefont {Chen}\ \emph
  {et~al.}(2019{\natexlab{a}})\citenamefont {Chen}, \citenamefont {Watanabe},
  \citenamefont {Yu}, \citenamefont {Guan},\ and\ \citenamefont {del
  Campo}}]{chen19an}%
  \BibitemOpen
  \bibfield  {author} {\bibinfo {author} {\bibfnamefont {Y.-Y.}\ \bibnamefont
  {Chen}}, \bibinfo {author} {\bibfnamefont {G.}~\bibnamefont {Watanabe}},
  \bibinfo {author} {\bibfnamefont {Y.-C.}\ \bibnamefont {Yu}}, \bibinfo
  {author} {\bibfnamefont {X.-W.}\ \bibnamefont {Guan}}, \ and\ \bibinfo
  {author} {\bibfnamefont {A.}~\bibnamefont {del Campo}},\ }\href {\doibase
  10.1038/s41534-019-0204-5} {\bibfield  {journal} {\bibinfo  {journal} {npj
  Quantum Information}\ }\textbf {\bibinfo {volume} {5}},\ \bibinfo {pages}
  {88} (\bibinfo {year} {2019}{\natexlab{a}})}\BibitemShut {NoStop}%
\bibitem [{\citenamefont {Hartmann}\ \emph {et~al.}(2019)\citenamefont
  {Hartmann}, \citenamefont {Mukherjee}, \citenamefont {Niedenzu},\ and\
  \citenamefont {Lechner}}]{hartmann19many}%
  \BibitemOpen
  \bibfield  {author} {\bibinfo {author} {\bibfnamefont {A.}~\bibnamefont
  {Hartmann}}, \bibinfo {author} {\bibfnamefont {V.}~\bibnamefont {Mukherjee}},
  \bibinfo {author} {\bibfnamefont {W.}~\bibnamefont {Niedenzu}}, \ and\
  \bibinfo {author} {\bibfnamefont {W.}~\bibnamefont {Lechner}},\ }\href
  {https://arxiv.org/abs/1912.08689} {\bibfield  {journal} {\bibinfo  {journal}
  {arXiv:1912.08689}\ } (\bibinfo {year} {2019})}\BibitemShut {NoStop}%
\bibitem [{\citenamefont {Revathy}\ \emph {et~al.}(2020)\citenamefont
  {Revathy}, \citenamefont {Mukherjee}, \citenamefont {Divakaran},\ and\
  \citenamefont {del Campo}}]{revathy20universal}%
  \BibitemOpen
  \bibfield  {author} {\bibinfo {author} {\bibfnamefont {B.~S.}\ \bibnamefont
  {Revathy}}, \bibinfo {author} {\bibfnamefont {V.}~\bibnamefont {Mukherjee}},
  \bibinfo {author} {\bibfnamefont {U.}~\bibnamefont {Divakaran}}, \ and\
  \bibinfo {author} {\bibfnamefont {A.}~\bibnamefont {del Campo}},\ }\href
  {https://arxiv.org/abs/2003.06607} {\bibfield  {journal} {\bibinfo  {journal}
  {arXiv:2003.06607}\ } (\bibinfo {year} {2020})}\BibitemShut {NoStop}%
\bibitem [{\citenamefont {Kerstjens}\ \emph {et~al.}(2018)\citenamefont
  {Kerstjens}, \citenamefont {Brown},\ and\ \citenamefont
  {Hovhannisyan}}]{alejandro}%
  \BibitemOpen
  \bibfield  {author} {\bibinfo {author} {\bibfnamefont {A.~P.}\ \bibnamefont
  {Kerstjens}}, \bibinfo {author} {\bibfnamefont {E.~G.}\ \bibnamefont
  {Brown}}, \ and\ \bibinfo {author} {\bibfnamefont {K.~V.}\ \bibnamefont
  {Hovhannisyan}},\ }\href {\doibase 10.1088/1367-2630/aaba02} {\bibfield
  {journal} {\bibinfo  {journal} {New Journal of Physics}\ }\textbf {\bibinfo
  {volume} {20}},\ \bibinfo {pages} {043034} (\bibinfo {year}
  {2018})}\BibitemShut {NoStop}%
\bibitem [{\citenamefont {Chen}\ \emph
  {et~al.}(2019{\natexlab{b}})\citenamefont {Chen}, \citenamefont {Sun},\ and\
  \citenamefont {Dong}}]{jinfuchen1}%
  \BibitemOpen
  \bibfield  {author} {\bibinfo {author} {\bibfnamefont {J.-F.}\ \bibnamefont
  {Chen}}, \bibinfo {author} {\bibfnamefont {C.-P.}\ \bibnamefont {Sun}}, \
  and\ \bibinfo {author} {\bibfnamefont {H.}~\bibnamefont {Dong}},\ }\href
  {\doibase https://doi.org/10.1103/PhysRevE.100.032144} {\bibfield  {journal}
  {\bibinfo  {journal} {Phys. Rev. E}\ }\textbf {\bibinfo {volume} {100}},\
  \bibinfo {pages} {032144} (\bibinfo {year} {2019}{\natexlab{b}})}\BibitemShut
  {NoStop}%
\bibitem [{\citenamefont {Chen}\ \emph
  {et~al.}(2019{\natexlab{c}})\citenamefont {Chen}, \citenamefont {Sun},\ and\
  \citenamefont {Dong}}]{jinfuchen2}%
  \BibitemOpen
  \bibfield  {author} {\bibinfo {author} {\bibfnamefont {J.-F.}\ \bibnamefont
  {Chen}}, \bibinfo {author} {\bibfnamefont {C.-P.}\ \bibnamefont {Sun}}, \
  and\ \bibinfo {author} {\bibfnamefont {H.}~\bibnamefont {Dong}},\ }\href
  {\doibase https://doi.org/10.1103/PhysRevE.100.062140} {\bibfield  {journal}
  {\bibinfo  {journal} {Phys. Rev. E}\ }\textbf {\bibinfo {volume} {100}},\
  \bibinfo {pages} {062140} (\bibinfo {year} {2019}{\natexlab{c}})}\BibitemShut
  {NoStop}%
\bibitem [{\citenamefont {Campaioli}\ \emph {et~al.}(2017)\citenamefont
  {Campaioli}, \citenamefont {Pollock}, \citenamefont {Binder}, \citenamefont
  {C\'eleri}, \citenamefont {Goold}, \citenamefont {Vinjanampathy},\ and\
  \citenamefont {Modi}}]{campaioli17enhancing}%
  \BibitemOpen
  \bibfield  {author} {\bibinfo {author} {\bibfnamefont {F.}~\bibnamefont
  {Campaioli}}, \bibinfo {author} {\bibfnamefont {F.~A.}\ \bibnamefont
  {Pollock}}, \bibinfo {author} {\bibfnamefont {F.~C.}\ \bibnamefont {Binder}},
  \bibinfo {author} {\bibfnamefont {L.}~\bibnamefont {C\'eleri}}, \bibinfo
  {author} {\bibfnamefont {J.}~\bibnamefont {Goold}}, \bibinfo {author}
  {\bibfnamefont {S.}~\bibnamefont {Vinjanampathy}}, \ and\ \bibinfo {author}
  {\bibfnamefont {K.}~\bibnamefont {Modi}},\ }\href {\doibase
  10.1103/PhysRevLett.118.150601} {\bibfield  {journal} {\bibinfo  {journal}
  {Phys. Rev. Lett.}\ }\textbf {\bibinfo {volume} {118}},\ \bibinfo {pages}
  {150601} (\bibinfo {year} {2017})}\BibitemShut {NoStop}%
\bibitem [{\citenamefont {Ferraro}\ \emph {et~al.}(2018)\citenamefont
  {Ferraro}, \citenamefont {Campisi}, \citenamefont {Andolina}, \citenamefont
  {Pellegrini},\ and\ \citenamefont {Polini}}]{marcello1}%
  \BibitemOpen
  \bibfield  {author} {\bibinfo {author} {\bibfnamefont {D.}~\bibnamefont
  {Ferraro}}, \bibinfo {author} {\bibfnamefont {M.}~\bibnamefont {Campisi}},
  \bibinfo {author} {\bibfnamefont {G.~M.}\ \bibnamefont {Andolina}}, \bibinfo
  {author} {\bibfnamefont {V.}~\bibnamefont {Pellegrini}}, \ and\ \bibinfo
  {author} {\bibfnamefont {M.}~\bibnamefont {Polini}},\ }\href {\doibase
  https://doi.org/10.1103/PhysRevLett.120.117702} {\bibfield  {journal}
  {\bibinfo  {journal} {Physical Review Letters}\ }\textbf {\bibinfo {volume}
  {120}},\ \bibinfo {pages} {117702} (\bibinfo {year} {2018})}\BibitemShut
  {NoStop}%
\bibitem [{\citenamefont {Andolina}\ \emph {et~al.}(2019)\citenamefont
  {Andolina}, \citenamefont {Keck}, \citenamefont {Mari}, \citenamefont
  {Campisi}, \citenamefont {Giovannetti},\ and\ \citenamefont
  {Polini}}]{marcello2}%
  \BibitemOpen
  \bibfield  {author} {\bibinfo {author} {\bibfnamefont {G.~M.}\ \bibnamefont
  {Andolina}}, \bibinfo {author} {\bibfnamefont {M.}~\bibnamefont {Keck}},
  \bibinfo {author} {\bibfnamefont {A.}~\bibnamefont {Mari}}, \bibinfo {author}
  {\bibfnamefont {M.}~\bibnamefont {Campisi}}, \bibinfo {author} {\bibfnamefont
  {V.}~\bibnamefont {Giovannetti}}, \ and\ \bibinfo {author} {\bibfnamefont
  {M.}~\bibnamefont {Polini}},\ }\href {\doibase
  https://doi.org/10.1103/PhysRevLett.122.047702} {\bibfield  {journal}
  {\bibinfo  {journal} {Physical Review Letters}\ }\textbf {\bibinfo {volume}
  {122}},\ \bibinfo {pages} {047702} (\bibinfo {year} {2019})}\BibitemShut
  {NoStop}%
\bibitem [{\citenamefont {Rossini}\ \emph {et~al.}(2019)\citenamefont
  {Rossini}, \citenamefont {Andolina}, \citenamefont {Rosa}, \citenamefont
  {Carrega},\ and\ \citenamefont {Polini}}]{marcello3}%
  \BibitemOpen
  \bibfield  {author} {\bibinfo {author} {\bibfnamefont {D.}~\bibnamefont
  {Rossini}}, \bibinfo {author} {\bibfnamefont {G.~M.}\ \bibnamefont
  {Andolina}}, \bibinfo {author} {\bibfnamefont {D.}~\bibnamefont {Rosa}},
  \bibinfo {author} {\bibfnamefont {M.}~\bibnamefont {Carrega}}, \ and\
  \bibinfo {author} {\bibfnamefont {M.}~\bibnamefont {Polini}},\ }\href
  {https://arxiv.org/abs/1912.07234} {\bibfield  {journal} {\bibinfo  {journal}
  {arXiv:1912.07234}\ } (\bibinfo {year} {2019})}\BibitemShut {NoStop}%
\bibitem [{\citenamefont {Correa}\ \emph {et~al.}(2015)\citenamefont {Correa},
  \citenamefont {Mehboudi}, \citenamefont {Adesso},\ and\ \citenamefont
  {Sanpera}}]{correa15individual}%
  \BibitemOpen
  \bibfield  {author} {\bibinfo {author} {\bibfnamefont {L.~A.}\ \bibnamefont
  {Correa}}, \bibinfo {author} {\bibfnamefont {M.}~\bibnamefont {Mehboudi}},
  \bibinfo {author} {\bibfnamefont {G.}~\bibnamefont {Adesso}}, \ and\ \bibinfo
  {author} {\bibfnamefont {A.}~\bibnamefont {Sanpera}},\ }\href {\doibase
  10.1103/PhysRevLett.114.220405} {\bibfield  {journal} {\bibinfo  {journal}
  {Phys. Rev. Lett.}\ }\textbf {\bibinfo {volume} {114}},\ \bibinfo {pages}
  {220405} (\bibinfo {year} {2015})}\BibitemShut {NoStop}%
\bibitem [{\citenamefont {Correa}\ \emph {et~al.}(2017)\citenamefont {Correa},
  \citenamefont {Perarnau-Llobet}, \citenamefont {Hovhannisyan}, \citenamefont
  {Hern\'{a}ndez-Santana}, \citenamefont {Mehboudi},\ and\ \citenamefont
  {Sanpera}}]{correa16low}%
  \BibitemOpen
  \bibfield  {author} {\bibinfo {author} {\bibfnamefont {L.~A.}\ \bibnamefont
  {Correa}}, \bibinfo {author} {\bibfnamefont {M.}~\bibnamefont
  {Perarnau-Llobet}}, \bibinfo {author} {\bibfnamefont {K.~V.}\ \bibnamefont
  {Hovhannisyan}}, \bibinfo {author} {\bibfnamefont {S.}~\bibnamefont
  {Hern\'{a}ndez-Santana}}, \bibinfo {author} {\bibfnamefont {M.}~\bibnamefont
  {Mehboudi}}, \ and\ \bibinfo {author} {\bibfnamefont {A.}~\bibnamefont
  {Sanpera}},\ }\href {\doibase 10.1103/PhysRevA.96.062103} {\bibfield
  {journal} {\bibinfo  {journal} {Phys. Rev. A}\ }\textbf {\bibinfo {volume}
  {96}},\ \bibinfo {pages} {062103} (\bibinfo {year} {2017})}\BibitemShut
  {NoStop}%
\bibitem [{\citenamefont {Kurizki}\ \emph {et~al.}(2017)\citenamefont
  {Kurizki}, \citenamefont {Alvarez},\ and\ \citenamefont
  {Zwick}}]{zwick17quantum}%
  \BibitemOpen
  \bibfield  {author} {\bibinfo {author} {\bibfnamefont {G.}~\bibnamefont
  {Kurizki}}, \bibinfo {author} {\bibfnamefont {G.~A.}\ \bibnamefont
  {Alvarez}}, \ and\ \bibinfo {author} {\bibfnamefont {A.}~\bibnamefont
  {Zwick}},\ }\href {\doibase https://doi.org/10.3390/technologies5010001}
  {\bibfield  {journal} {\bibinfo  {journal} {Technologies}\ }\textbf {\bibinfo
  {volume} {5}} (\bibinfo {year} {2017}),\
  https://doi.org/10.3390/technologies5010001}\BibitemShut {NoStop}%
\bibitem [{\citenamefont {Mukherjee}\ \emph {et~al.}(2019)\citenamefont
  {Mukherjee}, \citenamefont {Zwick}, \citenamefont {Ghosh}, \citenamefont
  {Chen},\ and\ \citenamefont {Kurizki}}]{mukherjee19enhanced}%
  \BibitemOpen
  \bibfield  {author} {\bibinfo {author} {\bibfnamefont {V.}~\bibnamefont
  {Mukherjee}}, \bibinfo {author} {\bibfnamefont {A.}~\bibnamefont {Zwick}},
  \bibinfo {author} {\bibfnamefont {A.}~\bibnamefont {Ghosh}}, \bibinfo
  {author} {\bibfnamefont {X.}~\bibnamefont {Chen}}, \ and\ \bibinfo {author}
  {\bibfnamefont {G.}~\bibnamefont {Kurizki}},\ }\href {\doibase
  10.1038/s42005-019-0265-y} {\bibfield  {journal} {\bibinfo  {journal} {Commun
  Phys}\ }\textbf {\bibinfo {volume} {2}},\ \bibinfo {pages} {162} (\bibinfo
  {year} {2019})}\BibitemShut {NoStop}%
\bibitem [{\citenamefont {Bhattacharjee}\ \emph {et~al.}(2020)\citenamefont
  {Bhattacharjee}, \citenamefont {Bhattacharya}, \citenamefont {Niedenzu},
  \citenamefont {Mukherjee},\ and\ \citenamefont
  {Dutta}}]{bhattacharjee20quantum}%
  \BibitemOpen
  \bibfield  {author} {\bibinfo {author} {\bibfnamefont {S.}~\bibnamefont
  {Bhattacharjee}}, \bibinfo {author} {\bibfnamefont {U.}~\bibnamefont
  {Bhattacharya}}, \bibinfo {author} {\bibfnamefont {W.}~\bibnamefont
  {Niedenzu}}, \bibinfo {author} {\bibfnamefont {V.}~\bibnamefont {Mukherjee}},
  \ and\ \bibinfo {author} {\bibfnamefont {A.}~\bibnamefont {Dutta}},\ }\href
  {\doibase 10.1088/1367-2630/ab61d6} {\bibfield  {journal} {\bibinfo
  {journal} {New Journal of Physics}\ }\textbf {\bibinfo {volume} {22}},\
  \bibinfo {pages} {013024} (\bibinfo {year} {2020})}\BibitemShut {NoStop}%
\bibitem [{\citenamefont {Erdman}\ \emph {et~al.}(2019)\citenamefont {Erdman},
  \citenamefont {Cavina}, \citenamefont {Fazio}, \citenamefont {Taddei},\ and\
  \citenamefont {Giovannetti}}]{erdman19maximum}%
  \BibitemOpen
  \bibfield  {author} {\bibinfo {author} {\bibfnamefont {P.~A.}\ \bibnamefont
  {Erdman}}, \bibinfo {author} {\bibfnamefont {V.}~\bibnamefont {Cavina}},
  \bibinfo {author} {\bibfnamefont {R.}~\bibnamefont {Fazio}}, \bibinfo
  {author} {\bibfnamefont {F.}~\bibnamefont {Taddei}}, \ and\ \bibinfo {author}
  {\bibfnamefont {V.}~\bibnamefont {Giovannetti}},\ }\href {\doibase
  10.1088/1367-2630/ab4dca} {\bibfield  {journal} {\bibinfo  {journal} {New
  Journal of Physics}\ }\textbf {\bibinfo {volume} {21}},\ \bibinfo {pages}
  {103049} (\bibinfo {year} {2019})}\BibitemShut {NoStop}%
\bibitem [{\citenamefont {Harrow}\ and\ \citenamefont
  {Montanaro}(2017)}]{harrow17quantum}%
  \BibitemOpen
  \bibfield  {author} {\bibinfo {author} {\bibfnamefont {A.~W.}\ \bibnamefont
  {Harrow}}\ and\ \bibinfo {author} {\bibfnamefont {A.}~\bibnamefont
  {Montanaro}},\ }\href {\doibase 10.1038/nature23458} {\bibfield  {journal}
  {\bibinfo  {journal} {Nat.}\ }\textbf {\bibinfo {volume} {549}},\ \bibinfo
  {pages} {203} (\bibinfo {year} {2017})}\BibitemShut {NoStop}%
\bibitem [{\citenamefont {Boixo}\ \emph {et~al.}(2018)\citenamefont {Boixo},
  \citenamefont {Isakov}, \citenamefont {Smelyanskiy}, \citenamefont {Babbush},
  \citenamefont {Ding}, \citenamefont {Jiang}, \citenamefont {Bremner},
  \citenamefont {Martinis},\ and\ \citenamefont
  {Neven}}]{boixo18characterizing}%
  \BibitemOpen
  \bibfield  {author} {\bibinfo {author} {\bibfnamefont {S.}~\bibnamefont
  {Boixo}}, \bibinfo {author} {\bibfnamefont {S.~V.}\ \bibnamefont {Isakov}},
  \bibinfo {author} {\bibfnamefont {V.~N.}\ \bibnamefont {Smelyanskiy}},
  \bibinfo {author} {\bibfnamefont {R.}~\bibnamefont {Babbush}}, \bibinfo
  {author} {\bibfnamefont {N.}~\bibnamefont {Ding}}, \bibinfo {author}
  {\bibfnamefont {Z.}~\bibnamefont {Jiang}}, \bibinfo {author} {\bibfnamefont
  {M.~J.}\ \bibnamefont {Bremner}}, \bibinfo {author} {\bibfnamefont {J.~M.}\
  \bibnamefont {Martinis}}, \ and\ \bibinfo {author} {\bibfnamefont
  {H.}~\bibnamefont {Neven}},\ }\href {\doibase 10.1038/s41567-018-0124-x}
  {\bibfield  {journal} {\bibinfo  {journal} {Nat. Phys.}\ }\textbf {\bibinfo
  {volume} {14}},\ \bibinfo {pages} {595} (\bibinfo {year} {2018})}\BibitemShut
  {NoStop}%
\bibitem [{\citenamefont {Watanabe}\ \emph {et~al.}(2017)\citenamefont
  {Watanabe}, \citenamefont {Venkatesh}, \citenamefont {Talkner},\ and\
  \citenamefont {del Campo}}]{watanabe17quantum}%
  \BibitemOpen
  \bibfield  {author} {\bibinfo {author} {\bibfnamefont {G.}~\bibnamefont
  {Watanabe}}, \bibinfo {author} {\bibfnamefont {B.~P.}\ \bibnamefont
  {Venkatesh}}, \bibinfo {author} {\bibfnamefont {P.}~\bibnamefont {Talkner}},
  \ and\ \bibinfo {author} {\bibfnamefont {A.}~\bibnamefont {del Campo}},\
  }\href {\doibase 10.1103/PhysRevLett.118.050601} {\bibfield  {journal}
  {\bibinfo  {journal} {Phys. Rev. Lett.}\ }\textbf {\bibinfo {volume} {118}},\
  \bibinfo {pages} {050601} (\bibinfo {year} {2017})}\BibitemShut {NoStop}%
\bibitem [{\citenamefont {Jaramillo}\ \emph {et~al.}(2016)\citenamefont
  {Jaramillo}, \citenamefont {Beau},\ and\ \citenamefont {del
  Campo}}]{campo16quantum}%
  \BibitemOpen
  \bibfield  {author} {\bibinfo {author} {\bibfnamefont {J.}~\bibnamefont
  {Jaramillo}}, \bibinfo {author} {\bibfnamefont {M.}~\bibnamefont {Beau}}, \
  and\ \bibinfo {author} {\bibfnamefont {A.}~\bibnamefont {del Campo}},\ }\href
  {\doibase https://doi.org/10.1088/1367-2630/18/7/075019} {\bibfield
  {journal} {\bibinfo  {journal} {New J. Phys.}\ }\textbf {\bibinfo {volume}
  {18}},\ \bibinfo {pages} {075019} (\bibinfo {year} {2016})}\BibitemShut
  {NoStop}%
\bibitem [{\citenamefont {Niedenzu}\ and\ \citenamefont
  {Kurizki}(2018)}]{niedenzu18cooperative}%
  \BibitemOpen
  \bibfield  {author} {\bibinfo {author} {\bibfnamefont {W.}~\bibnamefont
  {Niedenzu}}\ and\ \bibinfo {author} {\bibfnamefont {G.}~\bibnamefont
  {Kurizki}},\ }\href {\doibase https://doi.org/10.1088/1367-2630/aaed55}
  {\bibfield  {journal} {\bibinfo  {journal} {New J. Phys.}\ }\textbf {\bibinfo
  {volume} {20}},\ \bibinfo {pages} {113038} (\bibinfo {year}
  {2018})}\BibitemShut {NoStop}%
\bibitem [{\citenamefont {Kloc}\ \emph {et~al.}(2019)\citenamefont {Kloc},
  \citenamefont {Cejnar},\ and\ \citenamefont {Schaller}}]{kloc19collective}%
  \BibitemOpen
  \bibfield  {author} {\bibinfo {author} {\bibfnamefont {M.}~\bibnamefont
  {Kloc}}, \bibinfo {author} {\bibfnamefont {P.}~\bibnamefont {Cejnar}}, \ and\
  \bibinfo {author} {\bibfnamefont {G.}~\bibnamefont {Schaller}},\ }\href
  {\doibase 10.1103/PhysRevE.100.042126} {\bibfield  {journal} {\bibinfo
  {journal} {Phys. Rev. E}\ }\textbf {\bibinfo {volume} {100}},\ \bibinfo
  {pages} {042126} (\bibinfo {year} {2019})}\BibitemShut {NoStop}%
\bibitem [{\citenamefont {Abiuso}\ and\ \citenamefont
  {Giovannetti}(2019)}]{abiuso19non}%
  \BibitemOpen
  \bibfield  {author} {\bibinfo {author} {\bibfnamefont {P.}~\bibnamefont
  {Abiuso}}\ and\ \bibinfo {author} {\bibfnamefont {V.}~\bibnamefont
  {Giovannetti}},\ }\href {\doibase 10.1103/PhysRevA.99.052106} {\bibfield
  {journal} {\bibinfo  {journal} {Phys. Rev. A}\ }\textbf {\bibinfo {volume}
  {99}},\ \bibinfo {pages} {052106} (\bibinfo {year} {2019})}\BibitemShut
  {NoStop}%
\bibitem [{\citenamefont {Mukherjee}\ \emph {et~al.}(2020)\citenamefont
  {Mukherjee}, \citenamefont {Kofman},\ and\ \citenamefont
  {Kurizki}}]{mukherjee20anti}%
  \BibitemOpen
  \bibfield  {author} {\bibinfo {author} {\bibfnamefont {V.}~\bibnamefont
  {Mukherjee}}, \bibinfo {author} {\bibfnamefont {A.~G.}\ \bibnamefont
  {Kofman}}, \ and\ \bibinfo {author} {\bibfnamefont {G.}~\bibnamefont
  {Kurizki}},\ }\href {\doibase 10.1038/s42005-019-0272-z} {\bibfield
  {journal} {\bibinfo  {journal} {Communications Physics}\ }\textbf {\bibinfo
  {volume} {3}},\ \bibinfo {pages} {8} (\bibinfo {year} {2020})}\BibitemShut
  {NoStop}%
\bibitem [{\citenamefont {Camati}\ \emph {et~al.}(2020)\citenamefont {Camati},
  \citenamefont {Santos},\ and\ \citenamefont {Serra}}]{camati20employing}%
  \BibitemOpen
  \bibfield  {author} {\bibinfo {author} {\bibfnamefont {P.~A.}\ \bibnamefont
  {Camati}}, \bibinfo {author} {\bibfnamefont {J.~F.~G.}\ \bibnamefont
  {Santos}}, \ and\ \bibinfo {author} {\bibfnamefont {R.~M.}\ \bibnamefont
  {Serra}},\ }\href@noop {} {\  (\bibinfo {year} {2020})},\ \Eprint
  {http://arxiv.org/abs/2002.02039} {arXiv:2002.02039 [quant-ph]} \BibitemShut
  {NoStop}%
\bibitem [{\citenamefont {Breuer}\ and\ \citenamefont
  {Petruccione}(2002)}]{breuer02}%
  \BibitemOpen
  \bibfield  {author} {\bibinfo {author} {\bibfnamefont {H.~P.}\ \bibnamefont
  {Breuer}}\ and\ \bibinfo {author} {\bibfnamefont {F.}~\bibnamefont
  {Petruccione}},\ }\href@noop {} {\emph {\bibinfo {title} {The Theory of Open
  Quantum Systems}}}\ (\bibinfo  {publisher} {Oxford University Press},\
  \bibinfo {year} {2002})\BibitemShut {NoStop}%
\bibitem [{\citenamefont {Uzdin}\ \emph {et~al.}(2015)\citenamefont {Uzdin},
  \citenamefont {Levy},\ and\ \citenamefont {Kosloff}}]{uzdin15equivalence}%
  \BibitemOpen
  \bibfield  {author} {\bibinfo {author} {\bibfnamefont {R.}~\bibnamefont
  {Uzdin}}, \bibinfo {author} {\bibfnamefont {A.}~\bibnamefont {Levy}}, \ and\
  \bibinfo {author} {\bibfnamefont {R.}~\bibnamefont {Kosloff}},\ }\href
  {\doibase 10.1103/PhysRevX.5.031044} {\bibfield  {journal} {\bibinfo
  {journal} {Phys. Rev. X}\ }\textbf {\bibinfo {volume} {5}},\ \bibinfo {pages}
  {031044} (\bibinfo {year} {2015})}\BibitemShut {NoStop}%
\bibitem [{\citenamefont {Chru\ifmmode \acute{s}\else
  \'{s}\fi{}ci\ifmmode~\acute{n}\else \'{n}\fi{}ski}\ and\ \citenamefont
  {Kossakowski}(2010)}]{chruscinski10non}%
  \BibitemOpen
  \bibfield  {author} {\bibinfo {author} {\bibfnamefont {D.}~\bibnamefont
  {Chru\ifmmode \acute{s}\else \'{s}\fi{}ci\ifmmode~\acute{n}\else
  \'{n}\fi{}ski}}\ and\ \bibinfo {author} {\bibfnamefont {A.}~\bibnamefont
  {Kossakowski}},\ }\href {\doibase 10.1103/PhysRevLett.104.070406} {\bibfield
  {journal} {\bibinfo  {journal} {Phys. Rev. Lett.}\ }\textbf {\bibinfo
  {volume} {104}},\ \bibinfo {pages} {070406} (\bibinfo {year}
  {2010})}\BibitemShut {NoStop}%
\bibitem [{\citenamefont {Gelbwaser-Klimovsky}\ \emph
  {et~al.}(2013{\natexlab{b}})\citenamefont {Gelbwaser-Klimovsky},
  \citenamefont {Erez}, \citenamefont {Alicki},\ and\ \citenamefont
  {Kurizki}}]{klimovsky13work}%
  \BibitemOpen
  \bibfield  {author} {\bibinfo {author} {\bibfnamefont {D.}~\bibnamefont
  {Gelbwaser-Klimovsky}}, \bibinfo {author} {\bibfnamefont {N.}~\bibnamefont
  {Erez}}, \bibinfo {author} {\bibfnamefont {R.}~\bibnamefont {Alicki}}, \ and\
  \bibinfo {author} {\bibfnamefont {G.}~\bibnamefont {Kurizki}},\ }\href
  {\doibase 10.1103/PhysRevA.88.022112} {\bibfield  {journal} {\bibinfo
  {journal} {Phys. Rev. A}\ }\textbf {\bibinfo {volume} {88}},\ \bibinfo
  {pages} {022112} (\bibinfo {year} {2013}{\natexlab{b}})}\BibitemShut
  {NoStop}%
\bibitem [{\citenamefont {Rivas}\ \emph {et~al.}(2014)\citenamefont {Rivas},
  \citenamefont {Huelga},\ and\ \citenamefont {Plenio}}]{rivas14quantum}%
  \BibitemOpen
  \bibfield  {author} {\bibinfo {author} {\bibfnamefont {{\'{A}}.}~\bibnamefont
  {Rivas}}, \bibinfo {author} {\bibfnamefont {S.~F.}\ \bibnamefont {Huelga}}, \
  and\ \bibinfo {author} {\bibfnamefont {M.~B.}\ \bibnamefont {Plenio}},\
  }\href {\doibase 10.1088/0034-4885/77/9/094001} {\bibfield  {journal}
  {\bibinfo  {journal} {Reports on Progress in Physics}\ }\textbf {\bibinfo
  {volume} {77}},\ \bibinfo {pages} {094001} (\bibinfo {year}
  {2014})}\BibitemShut {NoStop}%
\bibitem [{\citenamefont {Katz}\ and\ \citenamefont
  {Kosloff}(2016)}]{gil16quantum}%
  \BibitemOpen
  \bibfield  {author} {\bibinfo {author} {\bibfnamefont {G.}~\bibnamefont
  {Katz}}\ and\ \bibinfo {author} {\bibfnamefont {R.}~\bibnamefont {Kosloff}},\
  }\href {\doibase https://doi.org/10.3390/e18050186} {\bibfield  {journal}
  {\bibinfo  {journal} {Entropy}\ }\textbf {\bibinfo {volume} {18}} (\bibinfo
  {year} {2016}),\ https://doi.org/10.3390/e18050186}\BibitemShut {NoStop}%
\bibitem [{\citenamefont {Nahar}\ and\ \citenamefont
  {Vinjanampathy}(2019)}]{nahar19preparations}%
  \BibitemOpen
  \bibfield  {author} {\bibinfo {author} {\bibfnamefont {S.}~\bibnamefont
  {Nahar}}\ and\ \bibinfo {author} {\bibfnamefont {S.}~\bibnamefont
  {Vinjanampathy}},\ }\href {\doibase 10.1103/PhysRevA.100.062120} {\bibfield
  {journal} {\bibinfo  {journal} {Phys. Rev. A}\ }\textbf {\bibinfo {volume}
  {100}},\ \bibinfo {pages} {062120} (\bibinfo {year} {2019})}\BibitemShut
  {NoStop}%
\bibitem [{\citenamefont {Mukherjee}\ \emph {et~al.}(2015)\citenamefont
  {Mukherjee}, \citenamefont {Giovannetti}, \citenamefont {Fazio},
  \citenamefont {Huelga}, \citenamefont {Calarco},\ and\ \citenamefont
  {Montangero}}]{mukherjee15efficiency}%
  \BibitemOpen
  \bibfield  {author} {\bibinfo {author} {\bibfnamefont {V.}~\bibnamefont
  {Mukherjee}}, \bibinfo {author} {\bibfnamefont {V.}~\bibnamefont
  {Giovannetti}}, \bibinfo {author} {\bibfnamefont {R.}~\bibnamefont {Fazio}},
  \bibinfo {author} {\bibfnamefont {S.~F.}\ \bibnamefont {Huelga}}, \bibinfo
  {author} {\bibfnamefont {T.}~\bibnamefont {Calarco}}, \ and\ \bibinfo
  {author} {\bibfnamefont {S.}~\bibnamefont {Montangero}},\ }\href {\doibase
  10.1088/1367-2630/17/6/063031} {\bibfield  {journal} {\bibinfo  {journal}
  {New Journal of Physics}\ }\textbf {\bibinfo {volume} {17}},\ \bibinfo
  {pages} {063031} (\bibinfo {year} {2015})}\BibitemShut {NoStop}%
\bibitem [{\citenamefont {Uzdin}\ \emph {et~al.}(2016)\citenamefont {Uzdin},
  \citenamefont {Levy},\ and\ \citenamefont {Kosloff}}]{uzdin16quantum}%
  \BibitemOpen
  \bibfield  {author} {\bibinfo {author} {\bibfnamefont {R.}~\bibnamefont
  {Uzdin}}, \bibinfo {author} {\bibfnamefont {A.}~\bibnamefont {Levy}}, \ and\
  \bibinfo {author} {\bibfnamefont {R.}~\bibnamefont {Kosloff}},\ }\href
  {\doibase 10.3390/e18040124} {\bibfield  {journal} {\bibinfo  {journal}
  {Entropy}\ }\textbf {\bibinfo {volume} {18}} (\bibinfo {year} {2016}),\
  10.3390/e18040124}\BibitemShut {NoStop}%
\bibitem [{\citenamefont {Thomas}\ \emph {et~al.}(2018)\citenamefont {Thomas},
  \citenamefont {Siddharth}, \citenamefont {Banerjee},\ and\ \citenamefont
  {Ghosh}}]{thomas18thermodynamics}%
  \BibitemOpen
  \bibfield  {author} {\bibinfo {author} {\bibfnamefont {G.}~\bibnamefont
  {Thomas}}, \bibinfo {author} {\bibfnamefont {N.}~\bibnamefont {Siddharth}},
  \bibinfo {author} {\bibfnamefont {S.}~\bibnamefont {Banerjee}}, \ and\
  \bibinfo {author} {\bibfnamefont {S.}~\bibnamefont {Ghosh}},\ }\href
  {\doibase 10.1103/PhysRevE.97.062108} {\bibfield  {journal} {\bibinfo
  {journal} {Phys. Rev. E}\ }\textbf {\bibinfo {volume} {97}},\ \bibinfo
  {pages} {062108} (\bibinfo {year} {2018})}\BibitemShut {NoStop}%
\bibitem [{\citenamefont {Pezzutto}\ \emph {et~al.}(2019)\citenamefont
  {Pezzutto}, \citenamefont {Paternostro},\ and\ \citenamefont
  {Omar}}]{pezzutto19an}%
  \BibitemOpen
  \bibfield  {author} {\bibinfo {author} {\bibfnamefont {M.}~\bibnamefont
  {Pezzutto}}, \bibinfo {author} {\bibfnamefont {M.}~\bibnamefont
  {Paternostro}}, \ and\ \bibinfo {author} {\bibfnamefont {Y.}~\bibnamefont
  {Omar}},\ }\href {\doibase 10.1088/2058-9565/aaf5b4} {\bibfield  {journal}
  {\bibinfo  {journal} {Quantum Science and Technology}\ }\textbf {\bibinfo
  {volume} {4}},\ \bibinfo {pages} {025002} (\bibinfo {year}
  {2019})}\BibitemShut {NoStop}%
\bibitem [{\citenamefont {Kofman}\ and\ \citenamefont
  {Kurizki}(2000)}]{kofman00acceleration}%
  \BibitemOpen
  \bibfield  {author} {\bibinfo {author} {\bibfnamefont {A.~G.}\ \bibnamefont
  {Kofman}}\ and\ \bibinfo {author} {\bibfnamefont {G.}~\bibnamefont
  {Kurizki}},\ }\href {\doibase https://doi.org/10.1038/35014537} {\bibfield
  {journal} {\bibinfo  {journal} {Nature}\ }\textbf {\bibinfo {volume} {405}},\
  \bibinfo {pages} {546} (\bibinfo {year} {2000})}\BibitemShut {NoStop}%
\bibitem [{\citenamefont {Kofman}\ and\ \citenamefont
  {Kurizki}(2001)}]{kofman01universal}%
  \BibitemOpen
  \bibfield  {author} {\bibinfo {author} {\bibfnamefont {A.~G.}\ \bibnamefont
  {Kofman}}\ and\ \bibinfo {author} {\bibfnamefont {G.}~\bibnamefont
  {Kurizki}},\ }\href {\doibase 10.1103/PhysRevLett.87.270405} {\bibfield
  {journal} {\bibinfo  {journal} {Phys. Rev. Lett.}\ }\textbf {\bibinfo
  {volume} {87}},\ \bibinfo {pages} {270405} (\bibinfo {year}
  {2001})}\BibitemShut {NoStop}%
\bibitem [{\citenamefont {Kofman}\ and\ \citenamefont
  {Kurizki}(2004)}]{kofman04unified}%
  \BibitemOpen
  \bibfield  {author} {\bibinfo {author} {\bibfnamefont {A.~G.}\ \bibnamefont
  {Kofman}}\ and\ \bibinfo {author} {\bibfnamefont {G.}~\bibnamefont
  {Kurizki}},\ }\href {\doibase 10.1103/PhysRevLett.93.130406} {\bibfield
  {journal} {\bibinfo  {journal} {Phys. Rev. Lett.}\ }\textbf {\bibinfo
  {volume} {93}},\ \bibinfo {pages} {130406} (\bibinfo {year}
  {2004})}\BibitemShut {NoStop}%
\bibitem [{\citenamefont {Erez}\ \emph {et~al.}(2008)\citenamefont {Erez},
  \citenamefont {Gordon}, \citenamefont {Nest},\ and\ \citenamefont
  {Kurizki}}]{erez08}%
  \BibitemOpen
  \bibfield  {author} {\bibinfo {author} {\bibfnamefont {N.}~\bibnamefont
  {Erez}}, \bibinfo {author} {\bibfnamefont {G.}~\bibnamefont {Gordon}},
  \bibinfo {author} {\bibfnamefont {M.}~\bibnamefont {Nest}}, \ and\ \bibinfo
  {author} {\bibfnamefont {G.}~\bibnamefont {Kurizki}},\ }\href {\doibase
  https://doi.org/10.1038/nature06873} {\bibfield  {journal} {\bibinfo
  {journal} {Nature}\ }\textbf {\bibinfo {volume} {452}},\ \bibinfo {pages}
  {724} (\bibinfo {year} {2008})}\BibitemShut {NoStop}%
\bibitem [{\citenamefont {\'Alvarez}\ \emph {et~al.}(2010)\citenamefont
  {\'Alvarez}, \citenamefont {Rao}, \citenamefont {Frydman},\ and\
  \citenamefont {Kurizki}}]{alvarez10zeno}%
  \BibitemOpen
  \bibfield  {author} {\bibinfo {author} {\bibfnamefont {G.~A.}\ \bibnamefont
  {\'Alvarez}}, \bibinfo {author} {\bibfnamefont {D.~D.~B.}\ \bibnamefont
  {Rao}}, \bibinfo {author} {\bibfnamefont {L.}~\bibnamefont {Frydman}}, \ and\
  \bibinfo {author} {\bibfnamefont {G.}~\bibnamefont {Kurizki}},\ }\href
  {\doibase 10.1103/PhysRevLett.105.160401} {\bibfield  {journal} {\bibinfo
  {journal} {Phys. Rev. Lett.}\ }\textbf {\bibinfo {volume} {105}},\ \bibinfo
  {pages} {160401} (\bibinfo {year} {2010})}\BibitemShut {NoStop}%
\bibitem [{\citenamefont {Almog}\ \emph {et~al.}(2011)\citenamefont {Almog},
  \citenamefont {Sagi}, \citenamefont {Gordon}, \citenamefont {Bensky},
  \citenamefont {Kurizki},\ and\ \citenamefont {Davidson}}]{almog11direct}%
  \BibitemOpen
  \bibfield  {author} {\bibinfo {author} {\bibfnamefont {I.}~\bibnamefont
  {Almog}}, \bibinfo {author} {\bibfnamefont {Y.}~\bibnamefont {Sagi}},
  \bibinfo {author} {\bibfnamefont {G.}~\bibnamefont {Gordon}}, \bibinfo
  {author} {\bibfnamefont {G.}~\bibnamefont {Bensky}}, \bibinfo {author}
  {\bibfnamefont {G.}~\bibnamefont {Kurizki}}, \ and\ \bibinfo {author}
  {\bibfnamefont {N.}~\bibnamefont {Davidson}},\ }\href {\doibase
  10.1088/0953-4075/44/15/154006} {\bibfield  {journal} {\bibinfo  {journal}
  {Journal of Physics B: Atomic, Molecular and Optical Physics}\ }\textbf
  {\bibinfo {volume} {44}},\ \bibinfo {pages} {154006} (\bibinfo {year}
  {2011})}\BibitemShut {NoStop}%
\bibitem [{\citenamefont {Mukherjee}\ \emph {et~al.}(2016)\citenamefont
  {Mukherjee}, \citenamefont {Niedenzu}, \citenamefont {Kofman},\ and\
  \citenamefont {Kurizki}}]{mukherjee16speed}%
  \BibitemOpen
  \bibfield  {author} {\bibinfo {author} {\bibfnamefont {V.}~\bibnamefont
  {Mukherjee}}, \bibinfo {author} {\bibfnamefont {W.}~\bibnamefont {Niedenzu}},
  \bibinfo {author} {\bibfnamefont {A.~G.}\ \bibnamefont {Kofman}}, \ and\
  \bibinfo {author} {\bibfnamefont {G.}~\bibnamefont {Kurizki}},\ }\href
  {\doibase 10.1103/PhysRevE.94.062109} {\bibfield  {journal} {\bibinfo
  {journal} {Phys. Rev. E}\ }\textbf {\bibinfo {volume} {94}},\ \bibinfo
  {pages} {062109} (\bibinfo {year} {2016})}\BibitemShut {NoStop}%
\bibitem [{\citenamefont {Rivas}\ and\ \citenamefont
  {Huelga}(2012)}]{rivas12open}%
  \BibitemOpen
  \bibfield  {author} {\bibinfo {author} {\bibfnamefont {A.}~\bibnamefont
  {Rivas}}\ and\ \bibinfo {author} {\bibfnamefont {S.~F.}\ \bibnamefont
  {Huelga}},\ }\href@noop {} {\emph {\bibinfo {title} {Open Quantum Systems}}}\
  (\bibinfo  {publisher} {Springer, Berlin, Heidelberg},\ \bibinfo {year} {2012})\BibitemShut
  {NoStop}%
\bibitem [{\citenamefont {Shahmoon}\ and\ \citenamefont
  {Kurizki}(2013)}]{shahmoon13engineering}%
  \BibitemOpen
  \bibfield  {author} {\bibinfo {author} {\bibfnamefont {E.}~\bibnamefont
  {Shahmoon}}\ and\ \bibinfo {author} {\bibfnamefont {G.}~\bibnamefont
  {Kurizki}},\ }\href {\doibase 10.1103/PhysRevA.87.013841} {\bibfield
  {journal} {\bibinfo  {journal} {Phys. Rev. A}\ }\textbf {\bibinfo {volume}
  {87}},\ \bibinfo {pages} {013841} (\bibinfo {year} {2013})}\BibitemShut
  {NoStop}%
\bibitem [{\citenamefont {Mukherjee}\ \emph {et~al.}(2013)\citenamefont
  {Mukherjee}, \citenamefont {Carlini}, \citenamefont {Mari}, \citenamefont
  {Caneva}, \citenamefont {Montangero}, \citenamefont {Calarco}, \citenamefont
  {Fazio},\ and\ \citenamefont {Giovannetti}}]{mukherjee13speeding}%
  \BibitemOpen
  \bibfield  {author} {\bibinfo {author} {\bibfnamefont {V.}~\bibnamefont
  {Mukherjee}}, \bibinfo {author} {\bibfnamefont {A.}~\bibnamefont {Carlini}},
  \bibinfo {author} {\bibfnamefont {A.}~\bibnamefont {Mari}}, \bibinfo {author}
  {\bibfnamefont {T.}~\bibnamefont {Caneva}}, \bibinfo {author} {\bibfnamefont
  {S.}~\bibnamefont {Montangero}}, \bibinfo {author} {\bibfnamefont
  {T.}~\bibnamefont {Calarco}}, \bibinfo {author} {\bibfnamefont
  {R.}~\bibnamefont {Fazio}}, \ and\ \bibinfo {author} {\bibfnamefont
  {V.}~\bibnamefont {Giovannetti}},\ }\href {\doibase
  10.1103/PhysRevA.88.062326} {\bibfield  {journal} {\bibinfo  {journal} {Phys.
  Rev. A}\ }\textbf {\bibinfo {volume} {88}},\ \bibinfo {pages} {062326}
  (\bibinfo {year} {2013})}\BibitemShut {NoStop}%
\bibitem [{\citenamefont {Misra}\ and\ \citenamefont
  {Sudarshan}(1977)}]{misra77the}%
  \BibitemOpen
  \bibfield  {author} {\bibinfo {author} {\bibfnamefont {B.}~\bibnamefont
  {Misra}}\ and\ \bibinfo {author} {\bibfnamefont {E.~C.~G.}\ \bibnamefont
  {Sudarshan}},\ }\href {\doibase 10.1063/1.523304} {\bibfield  {journal}
  {\bibinfo  {journal} {Journal of Mathematical Physics}\ }\textbf {\bibinfo
  {volume} {18}},\ \bibinfo {pages} {756} (\bibinfo {year} {1977})}\BibitemShut
  {NoStop}%
\bibitem [{\citenamefont {Itano}\ \emph {et~al.}(1990)\citenamefont {Itano},
  \citenamefont {Heinzen}, \citenamefont {Bollinger},\ and\ \citenamefont
  {Wineland}}]{itano90quantum}%
  \BibitemOpen
  \bibfield  {author} {\bibinfo {author} {\bibfnamefont {W.~M.}\ \bibnamefont
  {Itano}}, \bibinfo {author} {\bibfnamefont {D.~J.}\ \bibnamefont {Heinzen}},
  \bibinfo {author} {\bibfnamefont {J.~J.}\ \bibnamefont {Bollinger}}, \ and\
  \bibinfo {author} {\bibfnamefont {D.~J.}\ \bibnamefont {Wineland}},\ }\href
  {\doibase 10.1103/PhysRevA.41.2295} {\bibfield  {journal} {\bibinfo
  {journal} {Phys. Rev. A}\ }\textbf {\bibinfo {volume} {41}},\ \bibinfo
  {pages} {2295} (\bibinfo {year} {1990})}\BibitemShut {NoStop}%
\bibitem [{\citenamefont {Kosloff}\ and\ \citenamefont
  {Levy}(2014)}]{kosloff14quantum}%
  \BibitemOpen
  \bibfield  {author} {\bibinfo {author} {\bibfnamefont {R.}~\bibnamefont
  {Kosloff}}\ and\ \bibinfo {author} {\bibfnamefont {A.}~\bibnamefont {Levy}},\
  }\href {\doibase 10.1146/annurev-physchem-040513-103724} {\bibfield
  {journal} {\bibinfo  {journal} {Annual Review of Physical Chemistry}\
  }\textbf {\bibinfo {volume} {65}},\ \bibinfo {pages} {365} (\bibinfo {year}
  {2014})}\BibitemShut {NoStop}%
\bibitem [{\citenamefont {Klaers}\ \emph {et~al.}(2017)\citenamefont {Klaers},
  \citenamefont {Faelt}, \citenamefont {Imamoglu},\ and\ \citenamefont
  {Togan}}]{klaers17squeezed}%
  \BibitemOpen
  \bibfield  {author} {\bibinfo {author} {\bibfnamefont {J.}~\bibnamefont
  {Klaers}}, \bibinfo {author} {\bibfnamefont {S.}~\bibnamefont {Faelt}},
  \bibinfo {author} {\bibfnamefont {A.}~\bibnamefont {Imamoglu}}, \ and\
  \bibinfo {author} {\bibfnamefont {E.}~\bibnamefont {Togan}},\ }\href
  {\doibase 10.1103/PhysRevX.7.031044} {\bibfield  {journal} {\bibinfo
  {journal} {Phys. Rev. X}\ }\textbf {\bibinfo {volume} {7}},\ \bibinfo {pages}
  {031044} (\bibinfo {year} {2017})}\BibitemShut {NoStop}%
\bibitem [{\citenamefont {Ro{\ss}nagel}\ \emph {et~al.}(2016)\citenamefont
  {Ro{\ss}nagel}, \citenamefont {Dawkins}, \citenamefont {Tolazzi},
  \citenamefont {Abah}, \citenamefont {Lutz}, \citenamefont {Schmidt-Kaler},\
  and\ \citenamefont {Singer}}]{rossnage16a}%
  \BibitemOpen
  \bibfield  {author} {\bibinfo {author} {\bibfnamefont {J.}~\bibnamefont
  {Ro{\ss}nagel}}, \bibinfo {author} {\bibfnamefont {S.~T.}\ \bibnamefont
  {Dawkins}}, \bibinfo {author} {\bibfnamefont {K.~N.}\ \bibnamefont
  {Tolazzi}}, \bibinfo {author} {\bibfnamefont {O.}~\bibnamefont {Abah}},
  \bibinfo {author} {\bibfnamefont {E.}~\bibnamefont {Lutz}}, \bibinfo {author}
  {\bibfnamefont {F.}~\bibnamefont {Schmidt-Kaler}}, \ and\ \bibinfo {author}
  {\bibfnamefont {K.}~\bibnamefont {Singer}},\ }\href {\doibase
  10.1126/science.aad6320} {\bibfield  {journal} {\bibinfo  {journal}
  {Science}\ }\textbf {\bibinfo {volume} {352}},\ \bibinfo {pages} {325}
  (\bibinfo {year} {2016})}\BibitemShut {NoStop}%
\end{thebibliography}
\end{document}